\documentclass{ar2e}

\usepackage{amssymb}  
\usepackage{graphicx}
\usepackage{feynarts}

\def\beq{\begin{equation}}
\def\eeq#1{\label{#1}\end{equation}}
\def\eeqn{\end{equation}}

\newenvironment{Eqnarray}%
   {\arraycolsep 0.14em\begin{eqnarray}}{\end{eqnarray}}
\def\beqa{\begin{Eqnarray}}
\def\eeqa#1{\label{#1}\end{Eqnarray}}
\def\eeqan{\end{Eqnarray}}

\let\bar=\overbar

\def\VEV#1{\left\langle{ #1} \right\rangle}
\def\bra#1{\left\langle{ #1} \right|}
\def\ket#1{\left| {#1} \right\rangle}

\def\lsim{\mathrel{\raise.3ex\hbox{$<$\kern-.75em\lower1ex\hbox{$\sim$}}}}
\def\gsim{\mathrel{\raise.3ex\hbox{$>$\kern-.75em\lower1ex\hbox{$\sim$}}}}

\def\del{\partial}
\def\Dslash{\not{\hbox{\kern-4pt $D$}}}
\def\dslash{\not{\hbox{\kern-2pt $\del$}}}

\def\Pl{{\mbox{\scriptsize Pl}}}

\def\ee{e^+e^-}

\def\msb{{\bar{\scriptsize M \kern -1pt S}}}

\def\drb{{\bar{\scriptsize D \kern -1pt R}}}

\makeatletter
\def\section{\@startsection{section}{0}{\z@}{5.5ex plus .5ex minus
 1.5ex}{2.3ex plus .2ex}{\large\bf}}
\def\subsection{\@startsection{subsection}{1}{\z@}{3.5ex plus .5ex minus
 1.5ex}{1.3ex plus .2ex}{\normalsize\bf}}
\def\subsubsection{\@startsection{subsubsection}{2}{\z@}{-3.5ex plus
-1ex minus  -.2ex}{2.3ex plus .2ex}{\normalsize\sl}}

\renewcommand{\@makecaption}[2]{%
   \vskip 10pt
   \setbox\@tempboxa\hbox{\small #1: #2}
   \ifdim \wd\@tempboxa >\hsize     
       \small #1: #2\par          
     \else                        
       \hbox to\hsize{\hfil\box\@tempboxa\hfil}
   \fi}

 \def\citenum#1{{\def\@cite##1##2{##1}\cite{#1}}}
 
\newcount\@tempcntc
\def\@citex[#1]#2{\if@filesw\immediate\write\@auxout{\string\citation{#2}}\fi
  \@tempcnta\z@\@tempcntb\m@ne\def\@citea{}\@cite{\@for\@citeb:=#2\do
    {\@ifundefined
       {b@\@citeb}{\@citeo\@tempcntb\m@ne\@citea\def\@citea{,}{\bf ?}\@warning
       {Citation `\@citeb' on page \thepage \space undefined}}%
    {\setbox\z@\hbox{\global\@tempcntc0\csname b@\@citeb\endcsname\relax}%
     \ifnum\@tempcntc=\z@ \@citeo\@tempcntb\m@ne
       \@citea\def\@citea{,}\hbox{\csname b@\@citeb\endcsname}%
     \else
      \advance\@tempcntb\@ne
      \ifnum\@tempcntb=\@tempcntc
      \else\advance\@tempcntb\m@ne\@citeo
      \@tempcnta\@tempcntc\@tempcntb\@tempcntc\fi\fi}}\@citeo}{#1}}
\def\@citeo{\ifnum\@tempcnta>\@tempcntb\else\@citea\def\@citea{,}%
  \ifnum\@tempcnta=\@tempcntb\the\@tempcnta\else
  {\advance\@tempcnta\@ne\ifnum\@tempcnta=\@tempcntb \else\def\@citea{--}\fi
    \advance\@tempcnta\m@ne\the\@tempcnta\@citea\the\@tempcntb}\fi\fi}
\makeatother

\begin{document}

\input psfig.sty

%
%

\setcounter{topnumber}{2}
\renewcommand\topfraction{1.0}
\setcounter{bottomnumber}{1}
\renewcommand\bottomfraction{.3}
\setcounter{totalnumber}{3}
\renewcommand\textfraction{.0}
\renewcommand\floatpagefraction{.5}
\setcounter{dbltopnumber}{2}
\renewcommand\dbltopfraction{.7}
\renewcommand\dblfloatpagefraction{.5}
\def\sigv{\langle\sigma v\rangle}
\setlength{\baselineskip}{14pt}
\renewcommand\baselinestretch{1}

\setlength{\oddsidemargin}{0.5in}
\setlength{\evensidemargin}{0.3in}

\def\lsim{\mathrel{\raise.3ex\hbox{$<$\kern-.75em\lower1ex\hbox{$\sim$}}}}
\def\gsim{\mathrel{\raise.3ex\hbox{$>$\kern-.75em\lower1ex\hbox{$\sim$}}}}
\newcommand{\be}{\begin{equation}}
\newcommand{\ba}{\begin{eqnarray}}
\newcommand{\ea}{\end{eqnarray}}
\newcommand\simgreater{\buildrel > \over \sim}
\newcommand\simless{\buildrel < \over \sim}
\newcommand{\bt}{\begin{tabular}}
\newcommand{\et}{\end{tabular}}
\newcommand{\MeV}{{\rm MeV}} 
\newcommand{\fm}{{\rm fm}} 


\jname{Annu. Rev. Nucl. Part. Sci.}
\jyear{(2008)}
\jvol{Vol. 58}
\ARinfo{}

\title{\vspace{-5mm}
       Strategies for Determining the Nature of Dark Matter
       \\ \rule[3mm]{\textwidth}{1mm} \vspace{-15mm} }


\author{
Dan Hooper
\affiliation{Theoretical Astrophysics, Fermi National Accelerator Laboratory, \\
             Batavia, IL 60510,  USA;
             email: {\tt dhooper@fnal.gov}}
Edward A. Baltz
\affiliation{KIPAC, SLAC MS29, 2575 Sand Hill Road, Menlo Park, CA 94025, USA;
             email: {\tt eabaltz@slac.stanford.edu}} 
 \vspace{-5mm} }

\begin{keywords}
dark matter
\vspace{-5mm}
\end{keywords}

\begin{abstract}

In this review, we discuss the role of the various experimental programs taking part in the broader effort to identify the particle nature of dark matter. In particular, we focus on electroweak scale dark matter particles and discuss a wide range of search strategies being carried out and developed to detect them. These efforts include direct detection experiments, which attempt to observe the elastic scattering of dark matter particles with nuclei, indirect detection experiments, which search for photons, antimatter and neutrinos produced as a result of dark matter annihilations, and collider searches for new TeV-scale physics. Each of these techniques could potentially provide a different and complementary set of information related to the mass, interactions and distribution of dark matter. Ultimately, it is hoped that these many different tools will be used together to conclusively identify the particle or particles that constitute the dark matter of our universe.

\end{abstract}

\maketitle

 \section{INTRODUCTION}

There exists a wide array of evidence in support of the conclusion that most of
the matter in our universe is non-luminous. This includes observations of the rotational speeds of galaxies~\cite{rotationcurves}, the orbital velocities of galaxies within
clusters~\cite{clusters}, gravitational lensing~\cite{lensing}, the cosmic
microwave background~\cite{wmap}, the light element abundances~\cite{bbn} and
large scale structure~\cite{lss}.  Despite these many observational indications
of dark matter, it is clear that it does not consist of baryonic material or other known forms of matter. For the time being, we remain ignorant of the particle identity of this substance.

In this review, we summarize some of the most promising strategies and
techniques being pursued to elucidate the nature of dark matter. These efforts
include direct detection experiments designed to observe the elastic scattering
of dark matter particles with nuclei, indirect detection experiments which hope
to detect the annihilation products of dark matter such gamma rays, neutrinos,
positrons, antiprotons, antideuterons, synchrotron radiation and X-rays, and
collider searches for dark matter and associated particles.

The material presented here is not intended to be an exhaustive summary of the
field of dark matter physics. Such reviews can be found
elsewhere~\cite{Bertone:2004pz}. In this article, we limit our discussion to
the case of candidate dark matter particles with electroweak scale masses and
couplings. Furthermore, for the sake of length, we do not discuss every
experimental approach being pursued, but instead focus on several of the most
promising direct, indirect and collider efforts, and on the interplay and
complementarity between these various programs.

We would like to emphasize that the detection of dark matter particles in any
one of the experimental channels discussed here will not alone be sufficient to
conclusively identify the nature of dark matter. The direct or indirect
detection of the dark matter particles making up our galaxy's halo is highly unlikely
to provide enough information to reveal the underlying physics (supersymmetry, etc.) behind these particles.  In contrast, collider experiments may identify a
long-lived, weakly interacting particle, but will not be able to test its
cosmological stability or abundance. Only by combining the information provided
by many different experimental approaches is the mystery of dark
matter's particle nature likely to be solved. Although the detection of dark matter in any one search channel would constitute a discovery of the utmost importance, it would almost certainly leave many important questions unanswered.

\section{THE WIMP HYPOTHESIS} 
        
In this review, we limit our discussion to dark matter candidates which are heavy, electrically neutral and weakly-interacting. This class of particles, collectively known as WIMPs, are particularly well motivated, especially when their mass and couplings are tied to the physics of the electroweak scale. Before we discuss the experimental techniques for detecting dark matter particles, we will briefly discuss some of the most compelling motivations for electroweak-scale dark matter.

The challenge of stabilizing the mass of the Higgs boson ({\it ie.}~the
hierarchy problem) leads us to expect new forms of matter to appear at or near
the electroweak scale. The nature of any physics beyond the Standard Model
which might appear at the TeV scale, however, is tightly constrained by the precision
electroweak measurements made at LEP. In particular, new discrete symmetries
are required of most phenomenologically viable models of TeV scale
physics~\cite{hier}. Such symmetries naturally lead to a stable particle or particles, which may potentially constitute the dark matter of our universe. 

A number of extensions of the Standard Model have been proposed which introduce new particle content at or near the electroweak scale, and which include a discrete symmetry of the form required to stabilize a potential dark matter candidate. The most well studied example is the lightest neutralino in supersymmetric models. Others examples include Kaluza-Klein hypercharge gauge bosons in models with universal extra dimensions~\cite{kkdm}, and the lightest T-parity odd particle in little Higgs theories~\cite{lh}.

Each of these candidates have similar masses and couplings, and thus will undergo
similar thermal histories in the early universe.  At high temperatures, WIMPs
are abundant, being freely created and annihilated in pairs.  As the universe
expands and the temperature drops below the WIMPs' production threshold, however, the
number density of these particles becomes rapidly suppressed. Ultimately, the
WIMPs will ``freeze out'' and remain as a thermal relic of the universe's hot
youth. The resulting density of WIMPs is given by:
\beq \Omega_{\chi}h^2 = {s_0\over \rho_c/h^2}
\left( {45\over \pi g_*}\right)^{1/2} { x_f \over m_\Pl }{1\over \VEV{\sigma
    v}}, \eeq{Omegachi} where $s_0$ is the current entropy density of the
universe, $\rho_c$ is the critical density, $h$ is the (scaled) Hubble
constant, $g_*$ is the effective number of relativistic degrees of freedom at
the time that the dark matter particle goes out of thermal equilibrium, $m_\Pl$
is the Planck mass, $x_f = m/T_f \approx 25$ is the inverse freeze-out
temperature in units of the WIMP mass, and $\VEV{\sigma v}$ is the thermal
average of the dark matter pair annihilation cross section times the relative
velocity.

In order for this process to yield a thermal
abundance of dark matter within the range measured by WMAP ($0.095 < \Omega h^2 <
0.129$)~\cite{wmap}, the thermally averaged annihilation cross section
is required to be $\sigv \approx 3 \times 10^{-26}$ cm$^3$/s (or alternatively, $\sigv \approx
0.9$ pb). Remarkably, this is quite similar to the value obtained for a generic
electroweak mass particle annihilating through the exchange of the electroweak
gauge or Higgs bosons. In particular, we notice that $\VEV{\sigma v} = \pi
\alpha^2/8m^2$ leads us to a WIMP mass on the order of $m \sim 100$~GeV.

We conclude that if a stable, weakly interacting, electroweak-scale particle
exists, then it is likely to be present in the universe today with an abundance
similar to the measured dark matter density. With this in mind, we focus our
dark matter search strategy on this particularly well motivated scenario in
which the dark matter particle has electroweak interactions and a mass near the electroweak scale.



\section{DIRECT	DETECTION}
         \label{sec:section5}

Experiments such as XENON~\cite{xenon}, CDMS~\cite{cdmssi,cdmssd},
ZEPLIN~\cite{zeplin}, Edelweiss~\cite{edelweiss}, CRESST~\cite{cresst},
WARP~\cite{warp} and COUPP~\cite{coupp} are designed to detect dark matter particles through their elastic
scattering with nuclei. This class of techniques is collectively known as
direct detection, in contrast to indirect detection efforts which attempt to
observe the annihilation products of dark matter particles. 

The role played by direct detection is important for a number of
reasons. Firstly, although collider experiments may be capable of detecting dark matter
particles, they will not be able to distinguish a cosmologically stable
WIMP from a long-lived but unstable particle. More generally speaking,
colliders will not inform us as to the cosmological abundance of a WIMP they
might observe. Furthermore, while the mass of the dark matter particle could
potentially be measured by a collider experiment such as the LHC, its couplings
are much more difficult to access in this way. Direct detection experiments, in contrast, provide a valuable probe of the dark matter's couplings to the Standard
Model. Finally, the uncertainties involved in direct detection are likely to be significantly smaller than in most indirect
detection channels. Whereas indirect detection rates rely critically on the
distribution of dark matter, especially in high density regions, and on other
astrophysical properties such as the galactic magnetic and radiation fields,
direct detection experiments rely only on the local dark matter density and
velocity distribution.

The density of dark matter in the local neighborhood is inferred by fitting observations to models of the 
galactic halo.  These observations including the rotational speed of stars at the solar circle 
and other locations, the total projected mass density 
(estimated by considering the motion of stars perpendicular 
to the galactic disk), peak-to-trough variations in 
the rotation curve ({\it ie.} the `flatness constraint'), and 
microlensing. Taken together, these constraints can be used to estimate the local halo density to 
lie between 4 $\times$ 10$^{-25}$ g/cm$^{-3}$ and 13 $\times$ 10$^{-25}$ 
g/cm$^3$ ($0.22-0.73\,$GeV/cm$^3$)~\cite{GGTurner}.  Limits on the density of MACHO microlensing objects
imply that at least 80\% of this is cold dark matter.   The velocity of the
WIMPs is expected be close to the galactic rotation velocity,  $230 \pm 20$ 
km/sec~\cite{Drukier}.

These observations, however, only constrain the dark matter density as averaged over scales larger than a kiloparsec or so. In contrast, the solar system moves a distance of $\sim$$10^{-3}$ parsecs relative to the dark matter halo each year. If dark matter is distributed in an inhomogeneous way over milliparsec scales ({\it ie.} as a collection of dense clumps and voids), then the density along the path of the Earth, as seen by direct detection experiments, could be much larger or smaller than is inferred by the rotational dynamics of our galaxy. 

Throughout most of our galaxy's halo, however, inhomogeneities in the small scale dark matter distribution are not anticipated to be large. The vast majority of the dark matter in the inner regions of our galaxy has been in place for $\sim$$10^{10}$ years; ample time for the destruction of clumps through tidal interactions. Using high-resolution simulations, Helmi, White and Springel find that the dark matter in the solar neighborhood is likely to consist of a superposition of hundreds of thousands of dark matter streams, collectively representing a very smooth and homogeneous distribution \cite{white}. That being said, if we happen to find our Solar System residing in a overdense clump or stream of dark matter, high direct detection rates could lead us to mistakenly infer an artificially large WIMP-nucleon elastic scattering cross section.

The nuclear physics involved in WIMP-nuclei elastic scattering also introduces uncertainties which may ultimately limit the accuracy to which the dark matter's couplings to the Standard Model can be measured. In many models, including many supersymmetric models, the WIMP-nucleon scattering cross section is dominated by the $t$-channel exchange of a Higgs boson.  The coupling of the Higgs boson to the proton receives its dominant
contributions from two sources, the coupling of the Higgs to gluons
through a heavy quark loop and the direct coupling of the Higgs to strange
quarks~\cite{jungman}.  That means that this coupling depends on the parameter
\begin{equation}
         f_{Ts} =  {\bra{p}  m_s \bar s s \ket{p} \over \bra{p}  H_{QCD}
                           \ket{p}} \ ,
\label{fTsdefin}
\end{equation}
that is, the fraction of the mass of the proton that arises from the mass of
the non-valence strange quarks in the proton wavefunction.  It has been known
for some time that there is significant uncertainty in this
quantity~\cite{KaplanNelson}, and several recent papers have pointed out the
uncertainty this introduces to calculations of the WIMP-nucleon elastic
scattering cross section~\cite{Bottino,EllisUpdate}.  In particular, in the
case of WIMPs which couple dominantly to the strange content of the nucleon,
this can lead to an uncertainty in the direct detection cross section of a
factor of 4 or even larger~\cite{Bottino}. It is possible that this uncertainty
could be reduced in the future through the use of lattice gauge
theory~\cite{UKQCD,Weise}.

The processes of WIMP-nuclei elastic scattering can be naturally divided into
spin-dependent and spin-independent contributions.  The spin-independent, or
coherent scattering, term is enhanced in WIMP-nucleus cross sections by factors
of $A^2$, making it advantageous to use targets consisting of heavy nuclei.
This enhancement is due to the fact that the WIMP wavelength is of order the
size of the nucleus, thus the scattering amplitudes on individual nucleons add
coherently.  The spin-dependent contribution, in contrast, couples to the spin
of the target nuclei and scales with $J (J+1)$.  Naively, this could be
considered a coherent subtraction of amplitudes of opposite signs of pairs of
nucleons.  As the current spin-dependent scattering constraints are not strong
enough to test many dark matter models, we devote our attention primarily to
the process of spin-independent scattering.

The spin-independent WIMP-nucleus elastic scattering cross section is given by:
\begin{equation}
\label{sig}
\sigma \approx \frac{4 m^2_{X} m^2_{T}}{\pi (m_{X}+m_T)^2} [Z f_p + (A-Z) f_n]^2,
\end{equation}
where $m_T$ is the mass of the target nucleus, $m_X$ is the WIMP's mass and $Z$
and $A$ are the atomic number and atomic mass of the nucleus.  $f_p$ and $f_n$
are the WIMP's couplings to protons and neutrons, given by:
\begin{equation}
f_{p,n}=\sum_{q=u,d,s} f^{(p,n)}_{T_q} a_q \frac{m_{p,n}}{m_q} + \frac{2}{27} f^{(p,n)}_{TG} \sum_{q=c,b,t} a_q  \frac{m_{p,n}}{m_q},
\label{feqn}
\end{equation}
where $a_q$ are the WIMP-quark couplings and $f^{(p,n)}_{T_q}$ denote the quark content of the nucleon.

The first term in Eq.~\ref{feqn} 
corresponds to interactions with the quarks in the target nuclei. In the case of neutralino dark matter, this can occur through either $t$-channel CP-even Higgs exchange, or $s$-channel squark exchange:
\vspace{-0.2cm}

\begin{feynartspicture}(222,254)(3,4.3)
\FADiagram{ }
\FAProp(5.,20.)(20.,10.)(0.,){/Straight}{0}
\FALabel(0.,18.0)[b]{$\chi^0$}
\FAProp(20.,-5.0)(20.,10.0)(0.,){/ScalarDash}{0}
\FAProp( 5.,-15.0)(20.,-5.0)(0.,){/Straight}{+1}
\FALabel(0.,-18.0)[b]{$q$}
\FALabel(25.,0.0)[b]{$H, h$}
\FAProp(20.,10.)(35.,20.)(0.,){/Straight}{0}
\FAProp(20.,-5.)(35.,-15.0)(0.,){/Straight}{+1}
\FALabel(40.,18.0)[b]{$\chi^0$}
\FALabel(40.,-18.)[b]{$q$}



\FAProp(70.,14.)(85.,3.)(0.,){/Straight}{0}
\FALabel(65.,12.0)[b]{$\chi^0$}

\FAProp(70.,-7.0)(85.,3.0)(0.,){/Straight}{+1}
\FALabel(65.,-9.0)[b]{$q$}

\FAProp(85.,3.)(105.,3.0)(0.,){/ScalarDash}{0}
\FALabel(95.,9.)[t]{$\tilde{q}$}

\FAProp(120.,14.)(105.,3.0)(0.,){/Straight}{0}
\FALabel(125.,12.)[b]{$\chi^0$}

\FAProp(120.,-7.)(105.,3.0)(0.,){/Straight}{-1}
\FALabel(125.,-9.)[b]{$q$}
\end{feynartspicture}

\vspace{-4.5cm}
\noindent
The second term corresponds to interactions with the gluons in the target through a loop diagram (a quark/squark loop in the case of supersymmetry). $f^{(p)}_{TG}$ is given by $1 -f^{(p)}_{T_u}-f^{(p)}_{T_d}-f^{(p)}_{T_s} 
\approx 0.84$, and analogously, $f^{(n)}_{TG} \approx 0.83$. 

Besides its mass, the only thing we need to know about the WIMP itself to calculate this cross section are its couplings to quarks, $a_q$. In the case of neutralino dark matter, the value of this coupling depends on many features of the supersymmetric spectrum. The contribution resulting from Higgs exchange depends on the neutralino composition, as well as the Higgs masses and couplings.

In the case of heavy squarks, small wino component and little mixing between the CP-even Higgs bosons ($\cos \alpha \approx 1$), neutralino-nuclei elastic scattering is dominated by $H$ exchange with strange and bottom quarks, leading to a neutralino-nucleon cross section approximately given by:
\begin{equation}
\sigma_{\chi N} \sim \frac{g^2_1 g^2_2 f_{\tilde{B}} f_{\tilde{H}} \,m^4_N}{4\pi m^2_W \cos^2 \beta \, m^4_H} \bigg(f_{T_s}+\frac{2}{27}f_{TG}\bigg)^2, \,\,\,\, (m_{\tilde{q}}\, \rm{large}, \cos \alpha \approx 1),
\label{case1}
\end{equation}
where $f_{\tilde{B}}$ and $f_{\tilde{H}}$ denote the bino and higgsino
fractions of the lightest neutralino and $\tan \beta$ is the ratio of the vevs
of the two Higgs doublets in the MSSM.  Note that the coupling involves the
product of $f_{\tilde{B}}$ and $f_{\tilde{H}}$.  Neutralinos that are purely
gaugino-like or purely higgsino-like have zero cross section with nuclei.  The
fundamental reason for this is that the relevant vertex is
gaugino-higgsino-Higgs.  

If the heavier of the two CP-even Higgs bosons is very
heavy and/or $\tan \beta$ is small, scattering with up-type quarks through light Higgs exchange can dominate:
\begin{equation}
\sigma_{\chi N} \sim \frac{g^2_1 g^2_2  f_{\tilde{B}} f_{\tilde{H}} \, m^4_N}{4\pi m^2_W \, m^4_h} \bigg(f_{T_u}+\frac{4}{27}f_{TG}\bigg)^2, \,\,\,\, (m_{\tilde{q}}, m_H\, \rm{large}, \cos \alpha \approx 1).
\label{case2}
\end{equation}
If $\tan \beta$ and $m_H$ are large and the squarks somewhat light, elastic scattering can instead be dominated by squark exchange:
\begin{equation}
\sigma_{\chi N} \sim \frac{g^2_1 g^2_2  f_{\tilde{B}} f_{\tilde{H}} \, m^4_N}{4\pi m^2_W \cos^2 \beta \, m^4_{\tilde{q}}} \bigg(f_{T_s}+\frac{2}{27}f_{TG}\bigg)^2, \,\,\,\, (\tilde{q}\,\, \rm{dominated}, \tan \beta \gg 1).
\label{case3}
\end{equation}
From these expressions~\cite{scatteraq}, it is clear that the direct detection of dark matter alone will not be very capable of revealing much about supersymmetry or the other underlying physics. There are a large number of degeneracies which can lead to a given value of the WIMP-nucleon cross section. Only by combining this information with collider and/or indirect detection data can one hope to infer the nature of the dark matter particle.

Currently, the strongest limits on the spin-independent WIMP-nucleon cross section come from the XENON~\cite{xenon} and CDMS~\cite{cdmssi} experiments, which have obtained an upper bound on the cross section of a $\sim$100~GeV WIMP at the $\sim$$10^{-7}$~pb level. These constraints, along with those of other experiments, are shown in Fig.~\ref{directcurrent}.

\begin{figure}
\begin{center}

\resizebox{8.5cm}{!}{\includegraphics{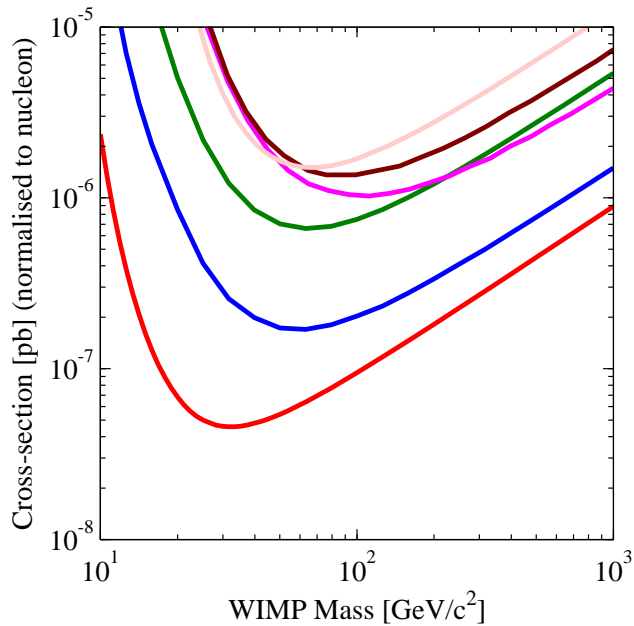}}
\caption{The current limits on the spin-independent WIMP-nuclei (normalized per nucleon) elastic scattering cross section, as a function of the WIMP mass. From bottom-to-top on the right side of the figure, these constraints come from the XENON~\cite{xenon}, CDMS~\cite{cdmssi}, WARP~\cite{warp}, ZEPLIN~\cite{zeplin}, Edelweiss~\cite{edelweiss} and CRESST~\cite{cresst} experiments. The plot was generated using the tool found at http://dmtools.berkeley.edu/limitplots/}
\label{directcurrent}
\end{center}
\end{figure}

There is currently a great deal of progress being made in the experimental
field of direct detection. Within the next several months (early 2008), the
CDMS collaboration is expected to release a new limit which will likely be the
most stringent (assuming no detection is made). In the meantime, the XENON
collaboration is preparing for a run with a larger detector, with results
expected within a year or so of this time. Beyond the next year or two, it
is difficult to foresee which experiment(s) will be leading this search. It is
still not clear whether detectors using liquid noble elements or cryogenic
technologies will advance most rapidly. For the time being, there are clear advantages to proceeding with multiple technologies.

Despite our inability to predict how this field will develop, it is reasonable to expect that by 2010 or so direct detection experiments will reach the $\sim 10^{-9}$ pb level of sensitivity. Roughly speaking, such cross sections are sufficient to test many, if not most, supersymmetric models, as well as many WIMPs candidates in other particle physics frameworks.

Given the rate at which direct dark matter experiments are developing, it is
interesting to recognize that such experiments are likely to see their first
evidence for WIMPs within the same time frame that the Large Hadron Collider
is expected to reveal the presence of the associated physics. In such a
scenario, it will be essential to compare the mass of the WIMP observed in each
experimental program. 

Direct detection experiments can determine the mass of the WIMP
by measuring the distribution of the recoil energy, $E_R$~\cite{Green:2007rb}.  This varies with the mass of the WIMP, with a 
resonance where the WIMP mass equals the target mass.  Roughly, one 
expects
\beq
        \VEV{E_R} \approx  { 2 v^2 m_T\over (1 + m_T/m_\chi)^2 }, 
\eeq{recoilenergy}
where $m_T$ is the target mass and $v$ is the WIMP velocity, 
with corrections depending on the precise target material
and the properties of
the detector~\cite{LewinSmith}.   Assuming the
standard velocity distribution in smooth halo models, 
with approximately 10\% uncertainty,
an experiment with a Xenon or Germanium target 
that detects 100 signal events for 
a WIMP of mass $m_X = 100$ GeV
can expect to measure the mass of this particle at the 20\% level, thus potentially confirming the cosmological stability (and abundance) of a WIMP detected at the LHC.  

If the WIMP mass inferred in a direct detection experiment was not consistent with that measured at the LHC, this could imply that different particle species are being observed, or could be the result of a nonstandard 
dark matter velocity distribution. In the future, directional dark matter detectors may help to clear up such a scenario.

\section{INDIRECT DETECTION}

In parallel to direct detection experiments, a wide range of indirect detection
programs have been developed to search for the annihilation products of dark
matter particles. In particular, searches are underway to detect
neutrinos from dark matter annihilations in the core of the Sun, antimatter
particles from dark matter annihilations in the galactic halo, and
photons from dark matter annihilations in the halo of the Milky Way,
galactic substructure and the dark matter distribution integrated over
cosmological volumes. In this section, we briefly describe the role of these
experimental programs in the overall strategy to reveal dark matter's identity.

\subsection{Gamma Rays}

Searches for prompt photons generated in dark matter annihilations have a key
advantage over other indirect detection channels in that they travel
essentially unimpeded from their production site.  In particular, gamma rays are not
deflected by magnetic fields, and thus can potentially provide valuable angular information. For example, point sources of
dark matter annihilation radiation might appear from high density regions such as the
Galactic Center or dwarf spheroidal galaxies.  Furthermore, over
galactic distance scales, gamma rays are not attenuated, and thus retain their
spectral information. In other words, the spectrum observed at Earth is the
same spectrum that was generated in the dark matter annihilations.

The spectrum of photons produced in dark matter annihilations depends on the
details of the WIMP being considered. Supersymmetric neutralinos, for example,
typically annihilate to final states consisting of heavy fermions and gauge or
Higgs bosons~\cite{jungman}.  Generally speaking, each of these annihilation
modes typically result in a very similar spectrum of gamma rays (see, however, Ref.~\cite{bringmann}). The gamma ray
spectrum from a WIMP which annihilates to light leptons can be quite different,
however. This can be particularly important in the case of Kaluza-Klein dark matter in
models with one universal extra dimension, for example, in which dark matter
particles annihilate significantly to $e^+ e^-$ and $\mu^+ \mu^-$~\cite{kkdm}.

%

The Galactic Center has long been considered to be one of the most promising
regions of the sky in which to search for gamma rays from dark matter
annihilations~\cite{gchist}. The prospects for this depend, however, on a
number of factors including the nature of the WIMP, the distribution of dark
matter in the region around the Galactic Center, and our understanding of the
astrophysical backgrounds.

The gamma ray spectrum produced through dark matter annihilations is given by
\begin{equation}
\Phi_{\gamma}(E_{\gamma},\psi) = \frac{1}{2}\sigv \frac{dN_{\gamma}}{dE_{\gamma}} \frac{1}{4\pi m^2_X} \int_{\rm{los}} \rho^2(r) dl(\psi) d\psi.
\label{flux1}
\end{equation}
Here,  $\sigv$ is the thermally averaged WIMP annihilation cross section, $m_X$ is the mass of the WIMP, $\psi$ is the
angle observed relative to the direction of the Galactic Center, $\rho(r)$ is
the dark matter density as a function of distance to the Galactic Center, and
the integral is performed over the line-of-sight. $dN_{\gamma}/dE_{\gamma}$ is
the gamma ray spectrum generated per WIMP annihilation.

Averaging over a solid angle centered around a direction, $\psi$, we arrive
at
\begin{equation}
\Phi_{\gamma}(E_{\gamma}) \approx 2.8 \times 10^{-12} \, \rm{cm}^{-2} \, \rm{s}^{-1} \, \frac{dN_{\gamma}}{dE_{\gamma}} \bigg(\frac{\sigv}{3 \times 10^{-26} \,\rm{cm}^3/\rm{s}}\bigg)  \bigg(\frac{1 \, \rm{TeV}}{m_{\rm{X}}}\bigg)^2 J(\Delta \Omega, \psi) \Delta \Omega,
\label{flux2}
\end{equation}
where $\Delta \Omega$ is the solid angle observed. The quantity $J(\Delta \Omega, \psi)$ depends only on the dark matter distribution, and is the average over the solid angle of the quantity:
\begin{equation}
J(\psi) = \frac{1}{8.5 \, \rm{kpc}} \bigg(\frac{1}{0.3 \, \rm{GeV}/\rm{cm}^3}\bigg)^2 \, \int_{\rm{los}} \rho^2(r(l,\psi)) dl.
\end{equation}
$J(\psi)$ is normalized such that a completely flat halo profile, with a density equal to the value at the solar circle, integrated along the line-of-sight to the Galactic Center would yield a value of one. In dark matter distributions favored by N-body simulations, however, this value can be much larger. The Narvarro-Frenk-White profile~\cite{nfw}, which is a commonly used benchmark halo model, leads to values of  $J(\Delta \Omega=10^{-5} \, {\rm sr}, \psi=0) \sim 10^5$. The effects adiabatic contraction due to the cooling of baryons is further expected to increase this quantity~\cite{ac}.

The recent discovery of a bright, very high-energy gamma ray source in the
galactic center region by the Atmospheric Cerenkov Telescopes HESS~\cite{hess},
MAGIC~\cite{magic}, WHIPPLE~\cite{whipple} and CANGAROO-II~\cite{cangaroo} has
made efforts to identify gamma rays from dark matter annihilations more
difficult. This source appears to be coincident with the dynamical center of
the Milky Way (Sgr A$^*$) and has no detectable angular extension (less than
1.2 arcminutes). Its spectrum is well described by a power-law,
$dN_{\gamma}/dE_{\gamma} \propto E_{\gamma}^{-\alpha}$, where $\alpha=2.25 \pm
0.04 (\rm{stat}) \pm 0.10 (\rm{syst})$ over the range of 160 GeV to 20
TeV. Although speculations were initially made that this source could be the
product of annihilations of very heavy ($\gsim 10$ TeV) dark matter
particles~\cite{actdark}, the spectral shape appears inconsistent with a dark
matter interpretation. The source of these gamma rays is more likely an
astrophysical accelerator associated with our Galaxy's central supermassive
black hole~\cite{hessastro}. Although this gamma ray source represents a formidable
background for GLAST and other experiments searching for dark matter
annihilation radiation~\cite{gabi}, it may be possible to reduce the impact of this and other backgrounds by studying the angular distribution of gamma rays from this region of the sky~\cite{Dodelson:2007gd}.

The prospects for identifying dark matter annihilation radiation from the Galactic Center depends critically on the unknown dark matter density within the inner parsecs of the Milky Way and on the properties of the astrophysical backgrounds present. If these characteristics are favorable, then the Galactic Center is very likely to be the most promising region of the sky to study. If not, other regions with high dark matter densities may be more advantageous.

Dwarf spheroidal galaxies within and near the Milky Way provide an opportunity to search for dark matter annihilation radiation with considerably less contamination from astrophysical backgrounds. The flux of gamma rays from dark matter annihilations in such objects, however, is also expected to be lower than from a cusp in the center of the Milky Way~\cite{Evans:2003sc,Bergstrom:2005qk,Strigari:2007at}. As a result, planned experiments are likely to observe dark matter annihilation radiation from dwarf galaxies only in the most favorable range of particle physics models. 

The integrated gamma ray signal from dark matter annihilations throughout the cosmological distribution of dark matter may also provide an opportunity to identify the products of dark matter annihilations. The ability of future gamma ray telescopes to identify a dark matter component of the diffuse flux depends strongly on the fraction of the extragalactic gamma ray background observed by the EGRET experiment which will be resolved as individual sources, such as blazars. If a large fraction of this background is resolved, the remaining extragalactic signal could potentially contain identifiable signatures of dark matter annihilations~\cite{egdiffuse}.

The telescopes potentially capable of detecting gamma rays from dark matter
annihilations in the near future include the satellite-based experiment
GLAST~\cite{glast}, and a number ground based Atmospheric Cerenkov Telescopes,
including HESS, MAGIC and VERITAS. The roles played by each of these two
classes of experiments in the search for dark matter are quite different. GLAST
will continuously observe a large fraction of the sky, but with an effective
area far smaller than possessed by ground based telescopes. Ground based
telescopes, in contrast, study the emission from a small angular field, but
with far greater exposure. Furthermore, while ground based telescopes can only
study gamma rays with energy greater than $\sim$100 GeV, GLAST will be able to
directly study gamma rays with energies over the range of 100 MeV to 300 GeV.

As a result of the different energy ranges accessible by these experiments,
searches for dark matter particles lighter than a few hundred GeV are most
promising with GLAST, while ground based telescopes are better suited for
heavier WIMPs. The large field-of-view of GLAST also makes it well suited for
measurements of the diffuse gamma ray background. GLAST is also expected to
detect a number of unidentified sources, some of which could potentially be signals of dark matter substructures. Follow up observations with ground based gamma ray
telescopes would be very useful for clarifying the nature of such sources.

\subsection{Antimatter}
\label{antimatter}
WIMP annihilations in the galactic halo generate charged anti-matter particles:
positrons, anti-protons and anti-deuterons. Unlike gamma rays, which travel
along straight lines, charged particles move under the influence of the
Galactic Magnetic Field, diffusing and losing energy, resulting in a diffuse
spectrum at Earth. By studying the cosmic anti-matter spectra, satellite-based
experiments such as PAMELA~\cite{pamela} and AMS-02~\cite{ams02} may be able to
identify signatures of dark matter. PAMELA began its three-year satellite mission in June
of 2006. AMS-02 is planned for later deployment onboard the International Space Station.

As compared to antiprotons and antideuterons, cosmic positrons are attractive probes of dark matter for several reasons. In particular, positrons lose the majority of their energy over typical length scales of a few kiloparsecs or less~\cite{baltzpos}. The cosmic positron spectrum, therefore, samples only the local dark matter distribution and is thus subject to considerably less uncertainty than the other anti-matter species. Additionally, data from the HEAT~\cite{heat} and AMS-01 experiments~\cite{ams01} contain features which could plausibly be the consequence of dark matter annihilations in the local halo.

The spectral shape of the cosmic positron spectrum generated in dark matter
annihilation depends on the leading annihilation modes of the WIMP in the low
velocity limit. Bino-like neutralinos, for example, typically annihilate to
heavy fermion pairs: $b\bar{b}$ with a small $\tau^+ \tau^-$ admixture, along
with a fraction to $t\bar{t}$ if $m_{\chi} \gsim m_t$. Wino or higgsino-like
neutralinos annihilate most efficiently to combinations of Higgs and gauge
bosons. In other particle dark matter candidates, such as Kaluza-Klein dark
matter in models with universal extra dimensions, annihilation to light charged
leptons can lead to a much harder positron spectrum than is expected from
neutralinos~\cite{kkpos}.

Once positrons are injected into the local halo through dark matter annihilations, they propagate under the influence of galactic magnetic fields, gradually losing energy through synchrotron emission and inverse Compton scattering with radiation fields, such as starlight and the cosmic microwave background. The spectrum observed at Earth is found by solving the diffusion-loss equation~\cite{diffusion}:
\begin{eqnarray}
\frac{\partial}{\partial t}\frac{dn_{e^{+}}}{dE_{e^{+}}} = \vec{\bigtriangledown} \cdot \bigg[K(E_{e^{+}},\vec{x})  \vec{\bigtriangledown} \frac{dn_{e^{+}}}{dE_{e^{+}}} \bigg] + \frac{\partial}{\partial E_{e^{+}}} \bigg[b(E_{e^{+}},\vec{x})\frac{dn_{e^{+}}}{dE_{e^{+}}}  \bigg] + Q(E_{e^{+}},\vec{x}),
\label{dif}
\end{eqnarray}
where $dn_{e^{+}}/dE_{e^{+}}$ is the number density of positrons per unit
energy, $K(E_{e^{+}},\vec{x})$ is the diffusion constant,
$b(E_{e^{+}},\vec{x})$ is the rate of energy loss and $Q(E_{e^{+}},\vec{x})$ is
the source term, which contains all of the information about the dark matter
annihilation modes, cross section, and distribution. To solve the
diffusion-loss equation, a set of boundary conditions must be adopted. In this
application, the boundary condition is described as the distance from the
galactic plane at which the positrons can freely escape, $L$. These diffusion
parameters can be constrained by studying the spectra of various species of
cosmic ray nuclei, most importantly the boron-to-carbon ratio \cite{btoc}.

\begin{figure}

\resizebox{6.5cm}{!}{\includegraphics{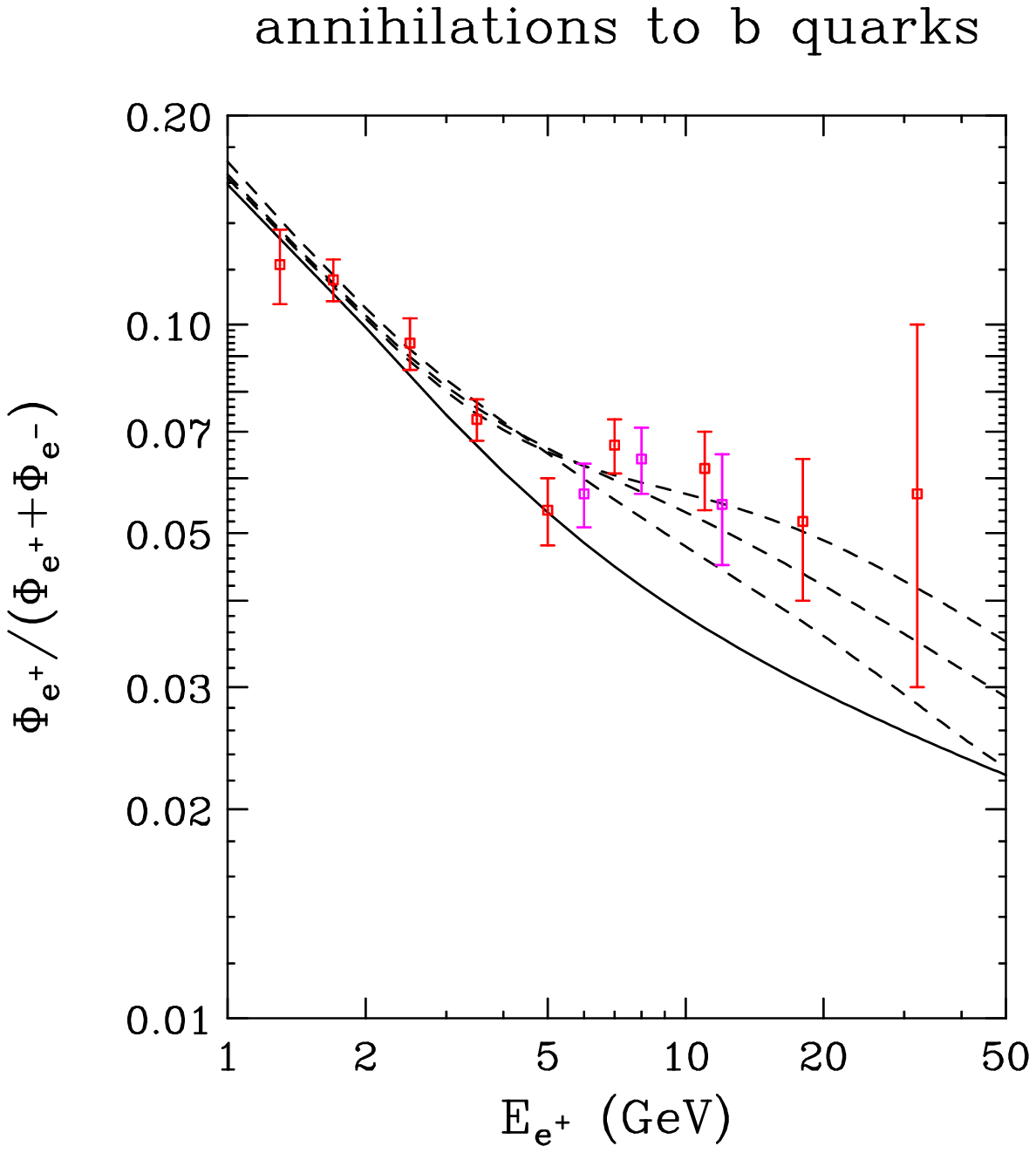}}
\resizebox{6.5cm}{!}{\includegraphics{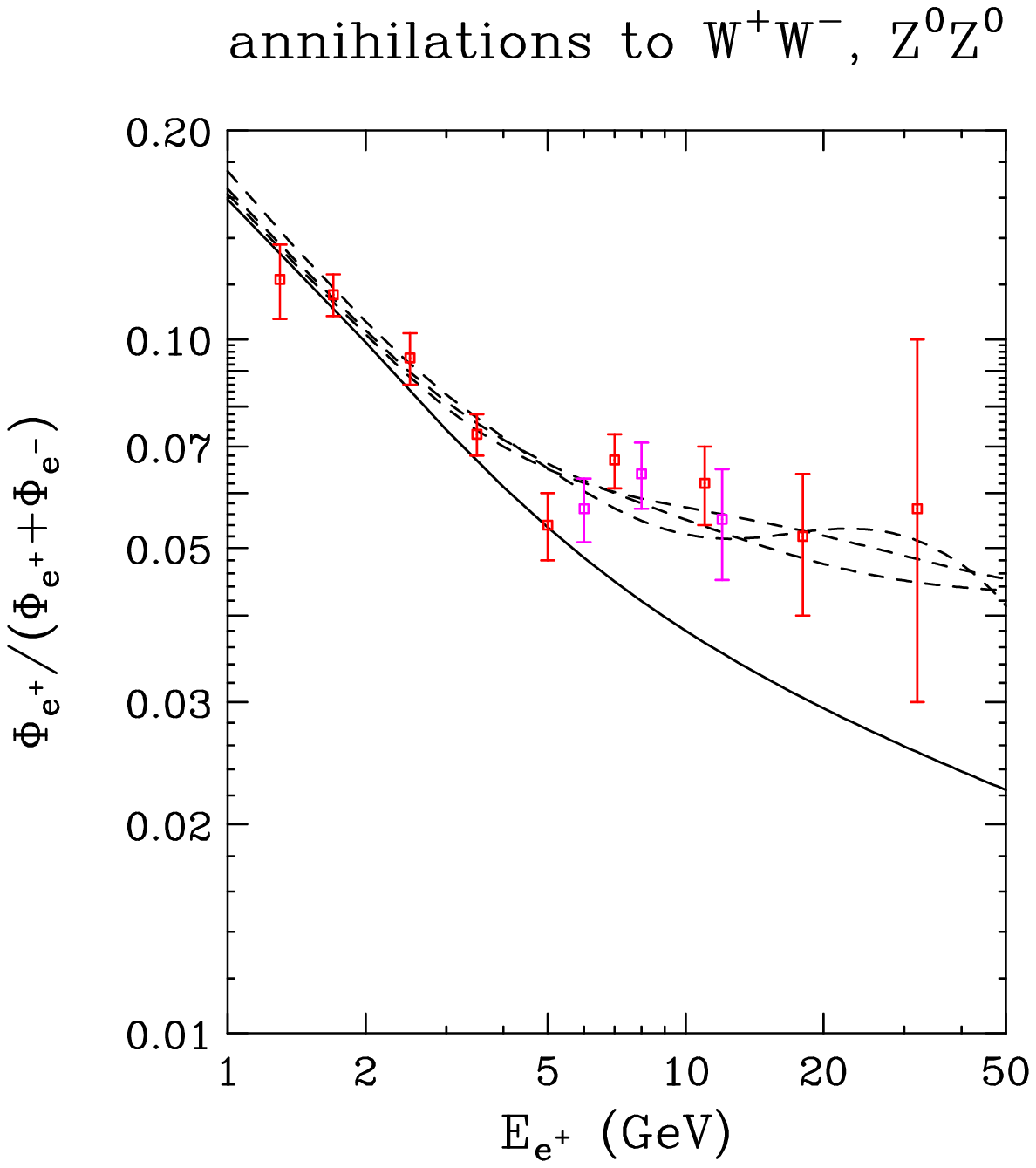}}
\caption{The positron fraction including the contribution from dark matter annihilations compared to the measurements of the HEAT experiment~\cite{heat}. Results are shown in each frame for WIMP masses of 100, 300 and 600 GeV. In the left (right) frame, the WIMP is assumed to annihilate to $b\bar{b}$ (a mixture of $ZZ$ and $W^+ W^-$). If an annihilation cross section of $\sigma v \approx 3\times 10^{-26}$ cm$^{3}$/s and a local density of 0.3 GeV/cm$^3$ is assumed, the annihilation rate must be boosted by a factor of approximately 50 or more to normalize to the HEAT data. The solid line shown denotes the prediction from the Galactic cosmic ray model of Ref.~\cite{secbg}.}\label{heat}
\end{figure}

In Fig.~\ref{heat}, the ratio of positrons to positrons plus electrons in the
cosmic ray spectrum is shown as a function of energy, including a possible
contribution from dark matter annihilations. Also shown are the measurements
from the HEAT experiment~\cite{heat}, which may possibly contain an excess in
comparison to standard astrophysical expectations at energies above 7 GeV or
so. While positrons from dark matter annihilations are indeed able to generate
this possible excess, it requires a somewhat larger annihilation rate than is
typically expected. In particular, if a smooth dark matter halo and an
annihilation cross section of $\sigma v \approx 3\times 10^{-26}$ cm$^{3}$/s
(as required to thermally produce the observed dark matter abundance via S-wave
processes) are assumed, the annihilation rate will be a factor of 50 or more
too low to generate the spectrum measured by HEAT. Fluctuations in the local
dark matter density, however, could lead to enhancements in the local
annihilation rate, known as the ``boost factor''. It is typically expected that
this quantity could be as large as 5 to 10. Although boost factors of 50 or
more are not impossible, such large values would be somewhat surprising.

If the positron flux observed by HEAT is in fact the result of annihilating dark matter, then the corresponding spectrum will be precisely measured by PAMELA~\cite{pamela} and AMS-02~\cite{ams02}. If not, then the detection of positrons from dark matter annihilations will be more difficult, but perhaps still possible~\cite{silkpos}.

Unlike gamma ray measurements of the Galactic Center or dwarf galaxies, observations of the cosmic positron spectrum (as well as the antiproton and antideuteron spectra) could potentially provide a measurement of the dark matter annihilation rate over large volumes of space. Such a measurement, therefore, could be used to determine the product of the WIMP's annihilation cross section and its density squared, averaged over the sampled volume (roughly a few cubic kiloparsec region, corresponding to the distance a typical positron travels from its point of origin before losing the majority of its energy). As a result of this limited range, only the dark matter distribution in the local halo is relevant to the observed cosmic positron flux. Assuming there are no very large and unknown clumps of dark matter in the surrounding kiloparsecs (which, although not impossible, is very unlikely~\cite{hoopertaylorsilk}), a measurement of the cosmic positron spectrum could be used to infer the dark matter particle's annihilation cross section (in the low velocity limit) with a comparatively modest degree of uncertainty coming from the unknown distribution of dark matter.

\subsection{Neutrino Telescopes}

Although dark matter annihilations in the galactic halo produce too few
neutrinos to be detected~\cite{neutrinoshalo}, annihilations which occur in the
center of the Sun could potentially generate an observable flux of high energy
neutrinos~\cite{neutrinosun}.

Dark matter particles scatter elastically with and become captured in the Sun at a rate given by~\cite{capture}
\begin{equation}
C^{\odot} \approx 3.35 \times 10^{19} \, \mathrm{s}^{-1} 
\left( \frac{\sigma_{\mathrm{H, SD}} +\, \sigma_{\mathrm{H, SI}}
+ 0.07 \, \sigma_{\mathrm{He, SI}}     } {10^{-7}\, \mathrm{pb}} \right)
\left( \frac{100 \, \mathrm{GeV}}{m_{X}} \right)^2 ,
\label{capture}
\end{equation}
where $m_{X}$ is the dark matter particle's mass. $\sigma_{\mathrm{H, SD}}$, $\sigma_{\mathrm{H, SI}}$ and $\sigma_{\mathrm{He, SI}}$ are the spin dependent (SD) and spin independent (SI) elastic scattering cross sections of the WIMP with hydrogen and helium nuclei, respectively. The factor of $0.07$ reflects the solar abundance of helium relative to hydrogen and well as dynamical factors and form factor suppression.

Notice that the capture rate is suppressed by two factors of the WIMP mass. One
of these is simply the result of the depleted number density of WIMPs in the
local halo ($n \propto 1/m$) while the second factor is the result of kinematic
suppression for the capture of a WIMP much heavier than the target nuclei, in
this case hydrogen or helium. If the WIMP's mass were comparable to the masses
of hydrogen or helium nuclei, these expressions would no longer be valid. For
WIMPs heavy enough to generate neutrinos detectable in the high-energy neutrino
telescopes, Eq.~\ref{capture} should be applicable. 

If the capture rate and annihilation cross sections are sufficiently large, equilibrium will be reached between these processes.  
For a number of WIMPs in the Sun, $N$, the rate of change of this
quantity is given by
\begin{equation}
\dot{N} = C^{\odot} - A^{\odot} N^2  ,
\end{equation}
where $C^{\odot}$ is the capture rate and $A^{\odot}$ is the 
annihilation cross section times the relative WIMP velocity per volume. $A^{\odot}$ can be approximated by
\begin{equation}
A^{\odot} \approx \frac{\sigv}{V_{\mathrm{eff}}}, 
\end{equation}
where $V_{\mathrm{eff}}$ is the effective volume of the core
of the Sun determined roughly by matching the core temperature with 
the gravitational potential energy of a single WIMP at the core
radius.  This was found in Refs.~\cite{equ1,equ2} to be
\begin{equation}
V_{\rm eff} \approx 5.7 \times 10^{27} \, \mathrm{cm}^3 
\left( \frac{100 \, \mathrm{GeV}}{m_{X}} \right)^{3/2} \;.
\end{equation}
The present WIMP annihilation rate in the Sun is given by
\begin{equation} 
\Gamma = \frac{1}{2} A^{\odot} N^2 = \frac{1}{2} \, C^{\odot} \, 
\tanh^2 \left( \sqrt{C^{\odot} A^{\odot}} \, t_{\odot} \right) \;, 
\end{equation}
where $t_{\odot} \approx 4.5$ billion years is the age of the solar system.
The annihilation rate is maximized when it reaches equilibrium with
the capture rate.  This occurs when 
\begin{equation}
\sqrt{C^{\odot} A^{\odot}} t_{\odot} \gg 1 \; .
\end{equation}
If this condition is met, the final annihilation rate (and corresponding neutrino flux and event rate) has no further dependence on the dark matter particle's annihilation cross section.

WIMPs can generate neutrinos through a wide range of annihilation channels. Annihilations to heavy quarks, tau leptons, gauge bosons and Higgs bosons can all generate neutrinos in the subsequent decay. In some models, WIMPs can also annihilate directly to neutrino pairs. 

Once produced, neutrinos can travel to the Earth where they can be detected.  The muon neutrino spectrum at the Earth from WIMP annihilations in the Sun is given by:
\begin{equation}
\frac{dN_{\nu_{\mu}}}{dE_{\nu_{\mu}}} = \frac{ C_{\odot} F_{\rm{Eq}}}{4 \pi D_{\rm{ES}}^2}   \bigg(\frac{dN_{\nu}}{dE_{\nu}}\bigg)^{\rm{Inj}},
\label{wimpflux}
\end{equation}
where $C_{\odot}$ is the WIMP capture rate in the Sun, $F_{\rm{Eq}}$ is the non-equilibrium suppression factor ($\approx 1$ for capture-annihilation equilibrium), $D_{\rm{ES}}$ is the Earth-Sun distance and $(\frac{dN_{\nu}}{dE_{\nu}})^{\rm{Inj}}$ is the neutrino spectrum from the Sun per WIMP annihilating. Due to $\nu_{\mu}-\nu_{\tau}$ vacuum oscillations, the muon neutrino flux from WIMP annihilations in the Sun observed at Earth is the average of the $\nu_{\mu}$ and $\nu_{\tau}$ components. 

Muon neutrinos produce muons in charged current interactions with ice or water nuclei inside or near the detector volume of a high energy neutrino telescope. The rate of neutrino-induced muons observed in a high-energy neutrino telescope is estimated by: 
\begin{equation}
N_{\rm{events}} \simeq \int \int \frac{dN_{\nu_{\mu}}}{dE_{\nu_{\mu}}}\, \frac{d\sigma_{\nu}}{dy}(E_{\nu_{\mu}},y) \,R_{\mu}((1-y)\,E_{\nu})\, A_{\rm{eff}} \, dE_{\nu_{\mu}} \, dy,
\end{equation}
where $\sigma_{\nu}(E_{\nu_{\mu}})$ is the neutrino-nucleon charged current interaction cross section, $(1-y)$ is the fraction of neutrino energy which goes into the muon, $A_{\rm{eff}}$ is the effective area of the detector, $R_{\mu}((1-y)\,E_{\nu})$ is the distance a muon of energy, $(1-y)\,E_{\nu}$, travels before falling below the muon energy threshold of the experiment (ranging from $\sim$1 to 100 GeV), called the muon range. 

The spectrum and flux of neutrinos generated in WIMP annihilations depends on
the annihilation modes which dominate, and thus is model dependent. For most annihilation modes, however, the variation from model-to-model is not dramatic. In Fig.~\ref{ratecompare},
the event rate in a kilometer-scale neutrino telescope (with a 50 GeV muon
energy threshold) is shown as a function of the WIMP's effective elastic
scattering cross section for a variety of annihilation
modes~\cite{Halzen:2005ar}. The effective elastic scattering cross section is
defined as $\sigma_{\rm{eff}} = \sigma_{\mathrm{H, SD}} +\, \sigma_{\mathrm{H,
    SI}} + 0.07 \, \sigma_{\mathrm{He, SI}}$, following
Eq.~\ref{capture}. These rates are indicative of that expected for experiments
such as IceCube at the South Pole~\cite{icecube}, or a future kilometer-scale
neutrino telescope built in the Mediterranean Sea~\cite{km3}. To detect
neutrinos from WIMP annihilations in the Sun over the background of atmospheric
neutrinos, a rate in the range of 10-100 events per square-kilometer, per year
is required.

\begin{figure}[t]
\includegraphics[width=2.2in,angle=90]{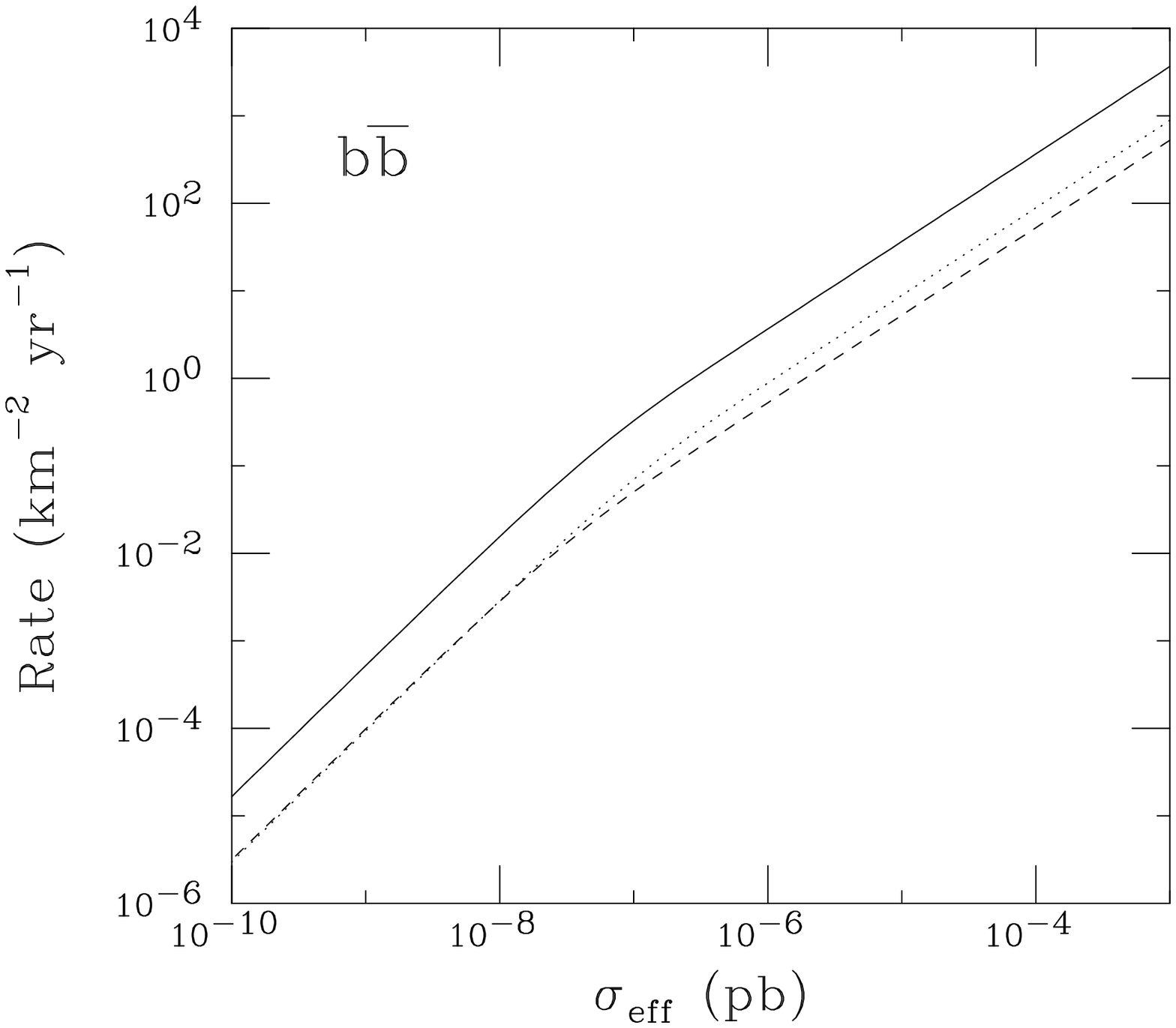}
\includegraphics[width=2.2in,angle=90]{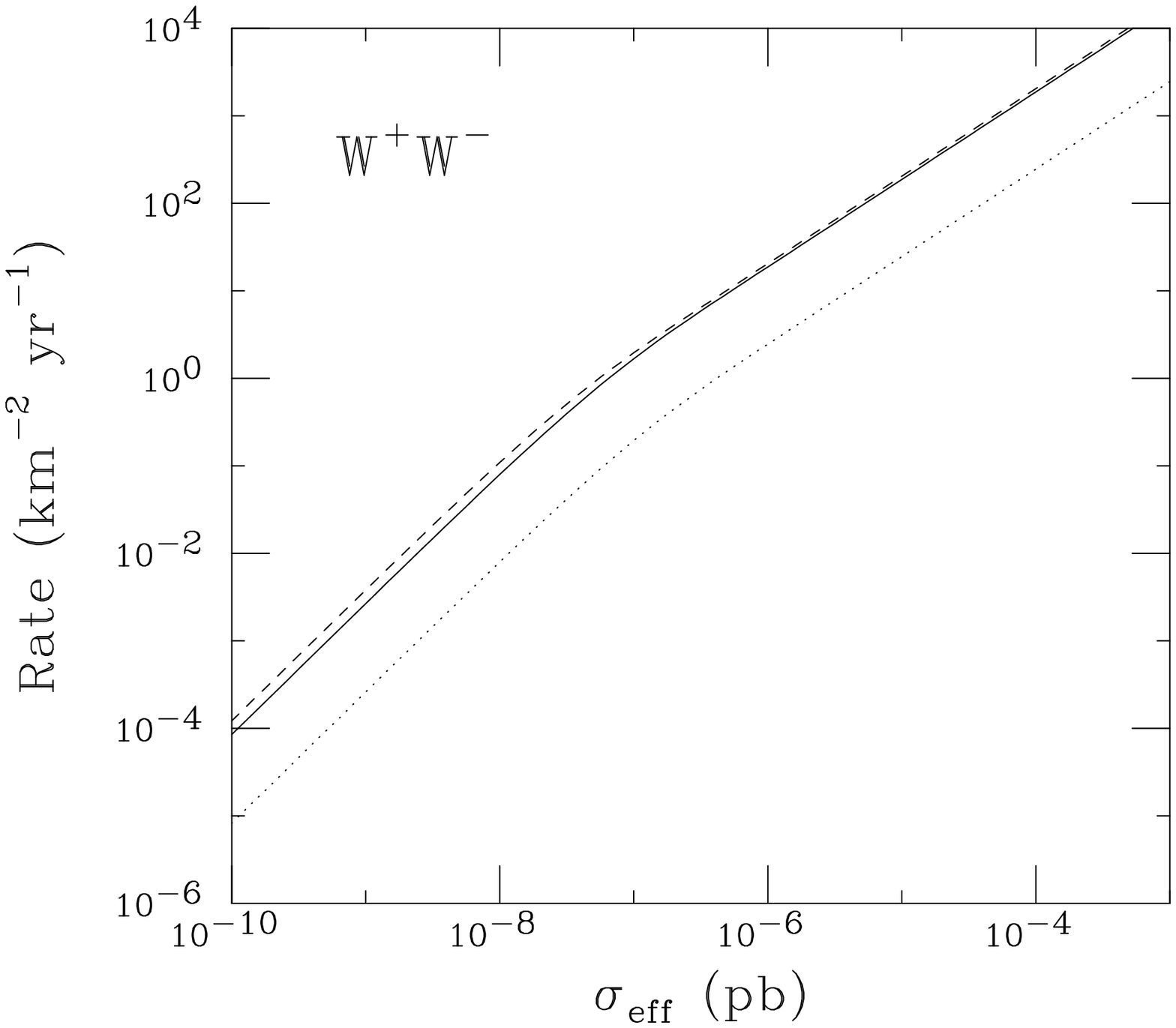}
 \\
\includegraphics[width=2.2in,angle=90]{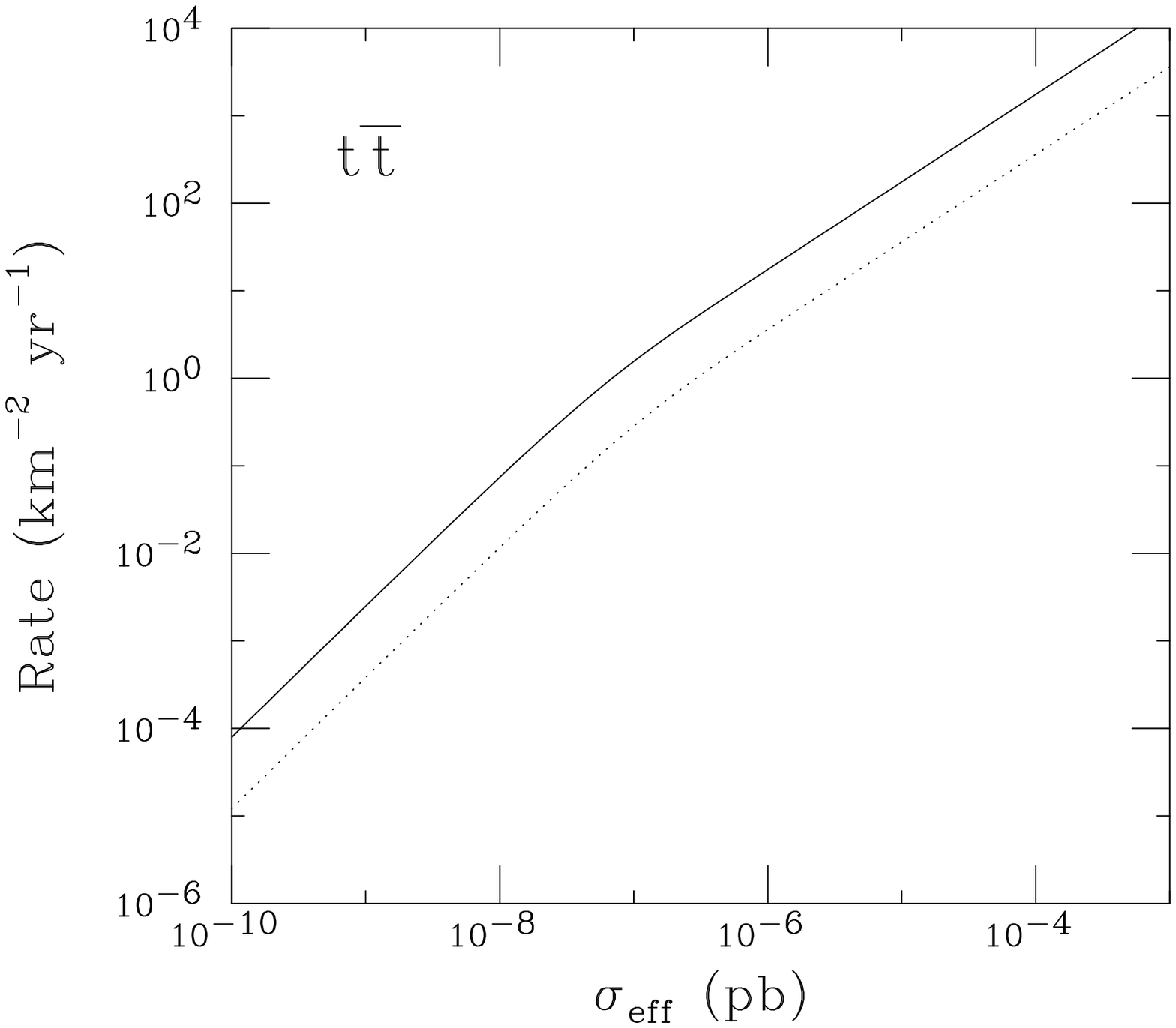}
\includegraphics[width=2.2in,angle=90]{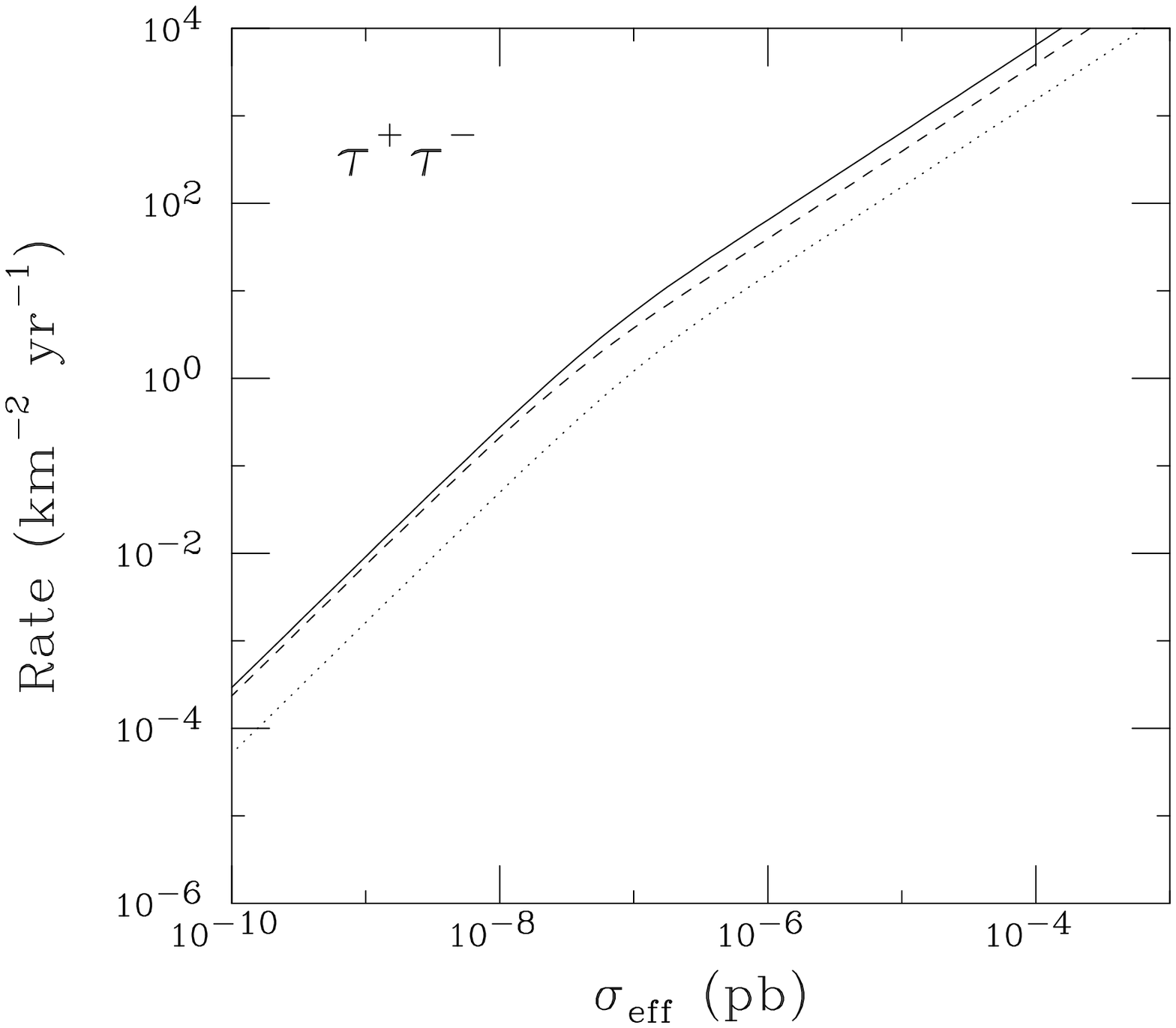}
 
\caption{The event rate in a kilometer-scale neutrino telescope such as IceCube as a function of the WIMP's effective elastic scattering cross section in the Sun for a variety of annihilation modes. The effective elastic scattering cross section is defined as $\sigma_{\rm{eff}} = \sigma_{\mathrm{H, SD}} +\, \sigma_{\mathrm{H, SI}}
+ 0.07 \, \sigma_{\mathrm{He, SI}}$, following Eq.~\ref{capture}. The dashes, solid and dotted lines correspond to WIMPs of mass 100, 300 and 1000 GeV, respectively.}
\label{ratecompare}
\end{figure}

Currently, the Super-Kamiokande experiment has placed the strongest bounds on high-energy neutrinos from the direction of the Sun~\cite{superk}. In this application, Super-K has two primary advantages over other neutrino detectors. Firstly, they have analyzed data over a longer period than most of their competitors, a total of nearly 1700 live days. Secondly, Super-K was designed to be sensitive to low energy ($\sim$GeV) neutrinos, which gives them an advantage in searching for lighter WIMPs. Super-K's upper limit on neutrino-induced muons above 1 GeV from WIMP annihilations in the Sun is approximately 1000 to 2000 per square kilometer per year for WIMPs heavier than 100 GeV, and approximately 2000 to 5000 per square kilometer per year for WIMPs in the 20 to 100 GeV range. The precise value of these limits depends on the WIMP annihilation mode considered.

The Amanda-II \cite{amanda}, Baksan \cite{baksan} and Macro \cite{macro} experiments have each placed limits on the flux of neutrino-induced muons from the Sun that are only slightly weaker than Super-Kamiokande's. The limit placed by the Amanda experiment resulted from only 144 live days of data. Having operated the detector for seven years, Amanda is expected to produce significantly improved bounds in the future.

In addition to these experiments, the next generation neutrino telescopes IceCube and Antares are currently under construction at the South Pole and in the Mediterranean, respectively. IceCube, with a full cubic kilometer of instrumented volume, will be considerably more sensitive to WIMP annihilations in the Sun than other planned or existing experiments \cite{icecube}. Antares, with less than one tenth of the effective area of IceCube, will have the advantage of a lower energy threshold, and may thus be more sensitive to low mass WIMPs \cite{antares}. Beyond Antares, there are also plans to build a kilometer-scale detector in the Mediterranean Sea~\cite{km3}.

From Fig.~\ref{ratecompare}, we see that a WIMP-proton elastic scattering cross section on the order of $10^{-6}$ pb or greater is needed if kilometer-scale neutrino telescopes are to detect a signal from dark matter annihilations. Elastic scattering cross sections of this size are constrained by the absence of a positive signal in direct detection experiments, however. Currently, the strongest constraints on the WIMP-nucleon, spin-independent elastic scattering cross section have been made by the XENON~\cite{xenon} and CDMS experiments~\cite{cdmssi}, who each place limits below $10^{-6}$ pb. Therefore, if current or planned neutrino telescopes are to detect neutrinos from dark matter annihilations in the Sun, they must scatter elastically with nuclei in the Sun via spin-dependent interactions, which are far less strongly constrained by direct detection experiments. The strongest bounds on the WIMP-proton spin-dependent cross section have been made by the NAIAD experiment~\cite{naiad}. This result limits the spin-dependent cross section with protons to be less than approximately 0.3 pb for a WIMP in the mass range of 50-100 GeV and less than 0.8 pb ($m_{X}$/500 GeV) for a heavier WIMP. The PICASSO~\cite{picasso} and CDMS~\cite{cdmssd} experiments have placed limits on the spin-dependent WIMP-proton cross section roughly one order of magnitude weaker than the NAIAD result. 

A WIMP with a largely spin-dependent scattering cross section with protons may thus be capable of generating large event rates in high energy neutrino telescopes. Considering, for example, a 300 GeV WIMP with an elastic scattering cross section near the experimental limit, Fig.~\ref{ratecompare} suggests that rates as high as $\sim 10^6$ per year could be generated if purely spin-dependent scattering contributes to the capture rate of WIMPs in the Sun.

The relative size of the spin-independent and spin-dependent elastic scattering cross sections depend on the nature of the WIMP in question. For a neutralino, these cross sections depend on its composition and on the mass spectrum of the exchanged Higgs bosons and squarks. Spin dependent, axial-vector, scattering of neutralinos with quarks within a nucleon is made possible through the t-channel exchange of a $Z$, or the s-channel exchange of a squark. Spin independent scattering occurs at the tree level through s-channel squark exchange and t-channel Higgs exchange, and at the one-loop level through diagrams involving a loop of quarks and/or squarks.

For higgsino-like or mixed higgsino-bino neutralinos, the spin dependent cross section can be somewhat larger than the spin independent, which is potentially well suited for the prospects for indirect detection. In particular, spin-dependent cross sections  as large as $\sim 10^{-3}$pb are possible even in models with very small spin-independent scattering rates. Such neutralinos would go easily undetected in all planned direct detection experiments, while still generating on the order of $\sim 1000$ events per year at IceCube.

\subsection{Synchrotron Emission}

As described in Sec.~\ref{antimatter}, electrons and positrons produced in dark matter annihilations travel under the influence of the Galactic Magnetic Field, losing energy through Compton scattering off of starlight, cosmic microwave background photons and far infrared emission from dust, and through synchrotron emission from interactions with the Galactic magnetic field. The relative importance of these processes depends on the energy densities of radiation and magnetic fields.

The processes of synchrotron emission and inverse Compton scattering each lead to potentially observable byproducts~\cite{syndm}. For dark matter particles with electroweak scale masses, the resulting synchrotron photons typically fall in the microwave frequency band, and thus are well suited for study with cosmic microwave background experiments~\cite{haze}. The inverse Compton scattering of highly relativistic electrons and positrons with starlight photons, on the other hand, can generate photons with MeV-GeV energies.

\section{THE ROLE OF COLLIDERS}

Among other new states, particles with TeV scale masses and QCD color are generic features of models of electroweak symmetry breaking. These particles appear as counterparts to the quarks to provide new physics associated with the 
generation of the large top quark mass.  In many scenarios, including supersymmetry, electroweak symmetry breaking arises as a result of 
radiative corrections
due to these particles, enhanced by the large coupling of the Higgs 
boson to the top quark. 

Any particle with these properties will be pair-produced at the Large Hadron Collider (LHC) with a cross section of tens of picobarns~\cite{sqgl}.  That
particle (or particles) will then subsequently decay to particles including quark or gluon jets and the lightest particle in the new sector ({\it ie.} the dark matter candidate) which proceeds to exit the detector unseen.  For any such model, the LHC experiments are, therefore, expected to observe large numbers of events with many hadronic jets and an imbalance
of measured momentum.  These `missing energy' events
are signatures of a wide range of models that contains an electroweak scale candidate for dark matter.

If TeV-scale supersymmetry exists in nature, it will very likely be within the discovery reach of the Large Hadron Collider (LHC). The rate of missing energy events depends strongly on the mass of the colored particles that are produced and only weakly on other properties of the model. In Fig.~\ref{fig:ATLAS}, the estimates of the ATLAS collaboration are shown for the discovery of missing
energy events~\cite{Tovey}. If squarks or gluinos have masses below 1~TeV, the missing energy events can be discovered with 
an integrated luminosity of 100 pb$^{-1}$, about 1\% of the LHC first-year
design luminosity.  Thus, we will know very early in the LHC program that
a WIMP candidate is being produced.

By studying the decays of squarks and/or gluinos it will also be possible to discover other superpartners at the LHC. For example, in many models, decays of the variety, $\tilde{q} \rightarrow \chi^0_2 q \rightarrow \tilde{l}^{\pm} l^{\mp} q \rightarrow  \chi^0_1 l^+ l^- q$, provide a clean signal of supersymmetry in the form of $l^+l^- +\, \rm{jets}\, + \,\,\rm{missing}\,\, E_T$. By studying the kinematics of these decays, the quantities $m_{\tilde{q}}$, $m_{\chi^0_2}$, $m_{\tilde{l}}$ and $m_{\chi^0_1}$ could each be potentially reconstructed~\cite{recon,drees,fitter}. More generally speaking, the LHC is, in most models, likely to measure the mass of the lightest neutralino to roughly 10\% accuracy, and may also be able to determine the masses of one or more of the other neutralinos, and any light sleptons~\cite{lhc}. Charginos are more difficult to study at the LHC.

\begin{figure}
\begin{center}

\resizebox{8.5cm}{!}{\includegraphics{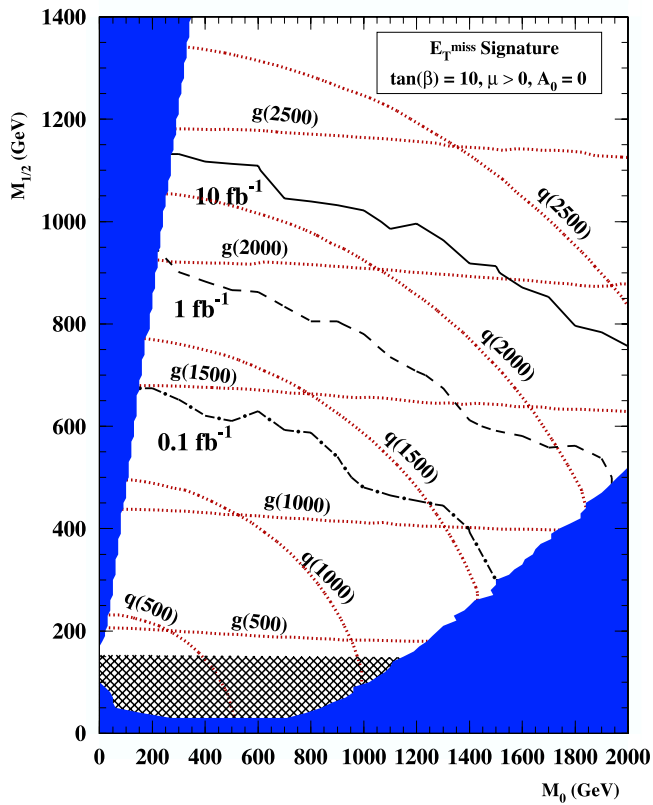}}
\caption{The discovery reach for supersymmetry via the missing energy plus jets signature by the ATLAS experiment at the LHC.
    The three sets of contours correspond to levels of integrated 
     luminosity at the LHC (in fb$^{-1}$), contours of constant squark
     mass, and contours of constant gluino mass.  From Ref.~\cite{Tovey}.}
\label{fig:ATLAS}
\end{center}
\end{figure}

The heavy, neutral Higgs bosons of the MSSM ($A$, $H$), can also be potentially produced and studied at the LHC. In particular, in models with large $\tan \beta$, heavy Higgs bosons have enhanced couplings to down-type fermions, thus leading to potentially observable di-tau final states. If enough of these events are observed, the masses of the heavy Higgs bosons could be potentially reconstructed, and $\tan \beta$ measured \cite{ditau,higgsmeasure}.

Prospects for the discovery of supersymmetry at the Tevatron, although not nearly as strong as at the LHC, are also exciting. The most likely discovery channel at the Tevatron is probably through clean tri-lepton plus missing energy events originating from the production of a chargino and a heavy neutralino, followed by a decay of the form, $\chi^{\pm} \chi^0_2 \rightarrow \tilde{\nu} l^{\pm} l^{+} \tilde{l^{-}} \rightarrow  \nu \chi^0_1 l^{\pm} l^+ l^- \chi^0_1$ \cite{fermilab}. Only models with rather light gauginos (neutralinos and charginos) and sleptons can be discovered in this way, however. For some of the recent results from supersymmetry searches at the Tevatron, see Ref.~\cite{recenttevatron}.

\begin{figure}

\resizebox{6.5cm}{!}{\includegraphics{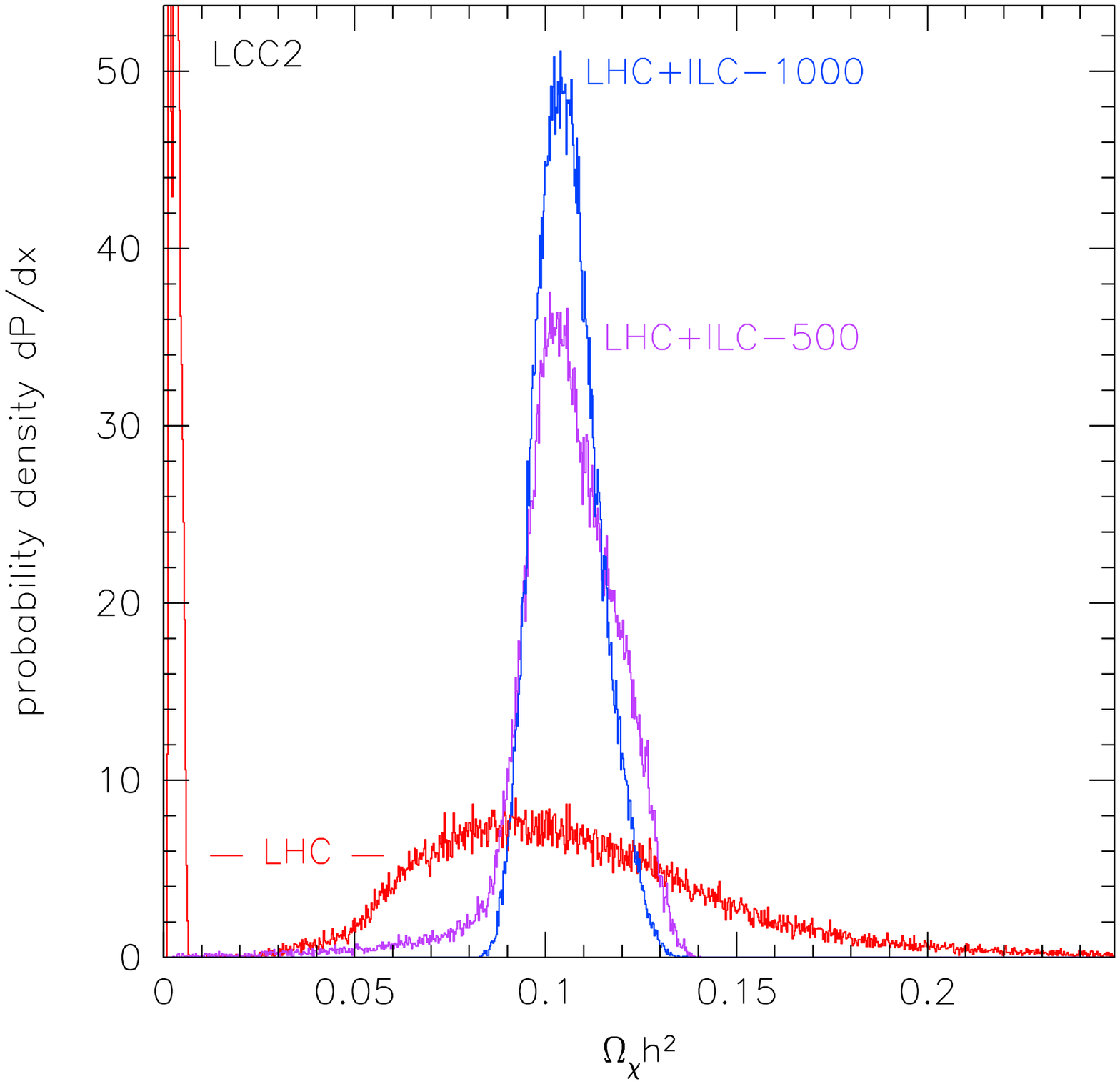}}
\resizebox{6.5cm}{!}{\includegraphics{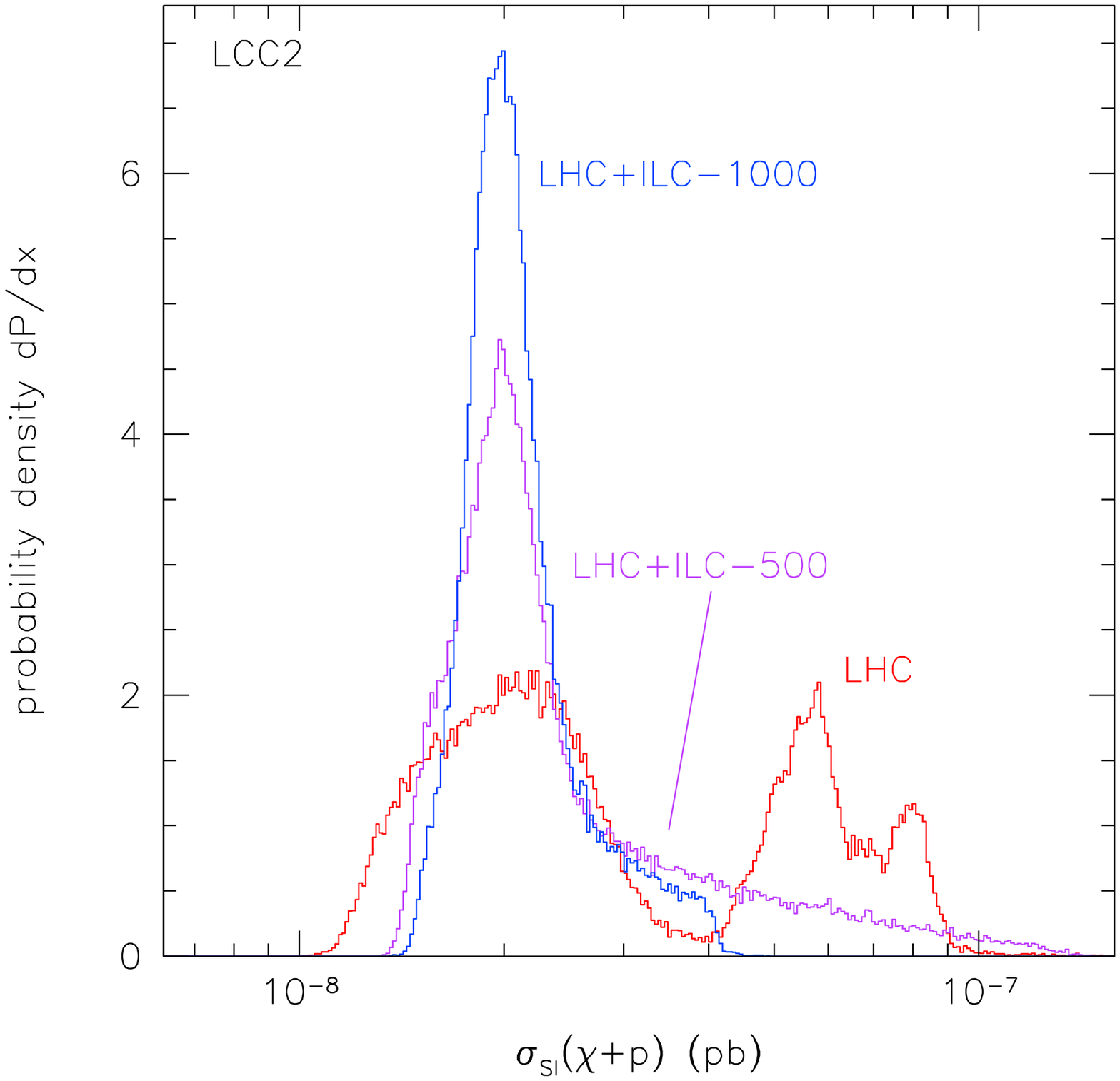}}
\caption{The ability of the LHC and a future ILC to infer the neutralino relic abundance (left) and elastic scattering cross section (right). The results shown are for a specific benchmark supersymmetry model (LCC2). See Ref.~\cite{Baltz:2006fm} for more details.}
\label{LCC2}
\end{figure}

Measurements of particle masses and other properties at the LHC can provide an essential cross check for direct and indirect detection channels. In particular, neither direct nor indirect detection experiments provide information capable of identifying the overall cosmological abundance of a WIMP, but instead infer only combinations of density and interaction cross section, leaving open the possibility that an observation may be generated by a sub-dominant component of the cosmological dark matter with a somewhat larger elastic scattering or annihilation cross section. Collider measurements can help to clarify this situation. 

In the left frame of Fig.~\ref{LCC2}, we show the ability of the LHC to infer the thermal neutralino relic abundance from measurements of sparticle masses and other properties. The results shown are for a specific benchmark supersymmetric model (see Ref.~\cite{Baltz:2006fm}), but are not atypical. In this case, the LHC can infer $\Omega_{\chi} h^2$ to lie roughly within 0.05 to 0.2 (assuming properties such as R-parity conservation), which, along with a detection in either a direct or indirect channel, would provide a strong confirmation that the observed neutralino does in fact constitute the bulk of the cosmological dark matter.

In the right frame of Fig.~\ref{LCC2}, the same supersymmetric model is considered, but instead showing the LHC's ability to determine the neutralino-nucleon elastic scattering cross section. This information would be very useful in combination with a direct detection signal. In particular, it would enable uncertainties in the local dark matter density to be reduced with confidence.
  
Also shown in each frame of Fig.~\ref{LCC2} are the results which could be obtained from a future 500 GeV or 1 TeV (center-of-mass) $e^+ e^-$ linear collider, such as the International Linear Collider (ILC)~\cite{ilc}. Such an experiment would have considerable advantages over hadron colliders such as the Tevatron or the LHC. Although hadron colliders can reach very high center-of-mass energies, and thus play an essential role as discovery machines, lepton colliders are best suited for lower energy, precision measurements. In particular, at an elec\-tron-pos\-i\-tron col\-lider, the process $\ee \to
X\bar X$ can provide an ex\-qui\-site diagnostic of the quantum numbers of the 
massive particle, $X$. As long as only the diagrams with annihilation 
through $\gamma$ and $Z$ are relevant, the angular distribution and
threshold shape of the reaction are characteristic for each spin, and the 
normalization of the cross section directly determines the $SU(2)\times U(1)$
quantum numbers.  These tests can be applied to any particles with electric
or weak charge whose pair-production thresholds lie in the range of the 
collider. Such a measurement could be used to pin down 
the spin and quantum numbers of a given particle and bring us a long way toward
the qualitative identification of the underlying model. 

We cannot emphasize enough the importance of the complementary roles played by each of the LHC and ILC programs. Whereas the LHC can more easily reach high energies and offers
very large cross sections for specific states of a model of new physics, the ILC will likely reach fewer states in the new particle
spectrum,  but will provide extremely incisive measurements of the properties of the particles that are
available to it.  Furthermore, the particles within the ILC reach are typically 
the ones on which the dark matter density depends most strongly.
Although both the LHC and ILC can make precision measurements, the 
measurements at the ILC typically have a more direct 
interpretation in terms of particle masses and couplings.

\section{SUMMARY}

In this review, we have attempted to summarize the diverse and complementary roles played by the various direct, indirect and collider searches for particle dark matter. As of 2008, there has not yet been a clear or conclusive detection of dark matter's non-gravitational interactions. There is reason to be optitmistic, however, that such a detection will be made within the next few years, moving the field beyond the discovery phase and into the measurement phase of the quest to reveal dark matter's nature and particle identity. As next generation direct detection experiments such as Super-CDMS, XENON-plus, LUX and others come online, most TeV-scale models containing a viable WIMP candidate will become within reach of these programs. Indirect detection experiments, including GLAST, VERITAS, HESS, MAGIC, PAMELA, AMS-02, IceCube and others are also rapidly advancing, and may see the first signals of dark matter annihilations. As the Large Hadron Collider begins its operation later this year, a new window into high-energy phenonoma will be opened. If dark matter is associated with physics of the electroweak scale, it is very likely to be within the discovery reach of this experiment.

The various experimental programs described in this review are each potentially capable of bringing very different measurments to the table. Although any one of these programs may be the first to discover particle dark matter, no single experiment or observation will answer all of our questions concerning this substance. Only by combining several of these detection methods together will it be possible to conclusively identify the dark matter of our universe.

\section*{Acknowledgments}
DH is supported by the United States Department of Energy and NASA grant NAG5-10842. Fermilab is operated by the Fermi Research Alliance, LLC under Contract No. DE-AC02-07CH11359 with the United States Department of Energy.




\begin{thebibliography}{99}


\bibitem{rotationcurves}
A.~Borriello and P.~Salucci, 
Mon.\ Not.\ Roy.\ Astron.\ Soc.\  {\bf 323}, 285 (2001) 
[arXiv:astro-ph/0001082].  
\bibitem{clusters}
F. Zwicky, Helv. Phys. Acta 6 (1933) 110.

\bibitem{lensing}
J.~A.~Tyson, G.~P.~Kochanski and I.~P.~Dell'Antonio,  
 Astrophys.\ J.\  {\bf 498}, L107 (1998)
[arXiv:astro-ph/9801193].
H.~Dahle, 
arXiv:astro-ph/0701598.
D.~Clowe, M.~Bradac, A.~H.~Gonzalez, M.~Markevitch, S.~W.~Randall, C.~Jones and D.~Zaritsky,
arXiv:astro-ph/0608407.

\bibitem{wmap}
  D.~N.~Spergel {\it et al.}  [WMAP Collaboration],
  Astrophys.\ J.\ Suppl.\  {\bf 170}, 377 (2007)
  [arXiv:astro-ph/0603449].

\bibitem{bbn}
K.~A.~Olive, G.~Steigman and T.~P.~Walker,
Phys.\ Rept.\  {\bf 333}, 389 (2000)
[arXiv:astro-ph/9905320].

\bibitem{lss}
 M.~Tegmark {\it et al.} 
[SDSS Collaboration],
Astrophys.\ J.\  {\bf 606}, 702 (2004)
[arXiv:astro-ph/0310725].


\bibitem{Bertone:2004pz}
  G.~Bertone, D.~Hooper and J.~Silk,
  Phys.\ Rept.\  {\bf 405}, 279 (2005)
  [arXiv:hep-ph/0404175].

\bibitem{hier}
  H.~C.~Cheng and I.~Low,
  JHEP {\bf 0309}, 051 (2003)
  [arXiv:hep-ph/0308199];
  J.~Wudka,
  arXiv:hep-ph/0307339.

\bibitem{kkdm}
G.~Servant and T.~M.~P.~Tait,
Nucl.\ Phys.\ B {\bf 650}, 391 (2003)
[arXiv:hep-ph/0206071];
H.~C.~Cheng, J.~L.~Feng and K.~T.~Matchev,
Phys.\ Rev.\ Lett.\  {\bf 89}, 211301 (2002)
[arXiv:hep-ph/0207125];
  D.~Hooper and S.~Profumo,
  Phys.~Rept., in press, arXiv:hep-ph/0701197.

\bibitem{lh}
  H.~C.~Cheng and I.~Low,
  JHEP {\bf 0309}, 051 (2003)
  [arXiv:hep-ph/0308199];
  H.~C.~Cheng and I.~Low,
  JHEP {\bf 0408}, 061 (2004)
  [arXiv:hep-ph/0405243].



\bibitem{xenon}
  J.~Angle {\it et al.}  [XENON Collaboration],
  arXiv:0706.0039 [astro-ph].

\bibitem{cdmssi}
 D.~S.~Akerib {\it et al.}  [CDMS Collaboration],
  arXiv:astro-ph/0509259.

\bibitem{cdmssd}
 D.~S.~Akerib {\it et al.}  [CDMS Collaboration],
  arXiv:astro-ph/0509269.


\bibitem{zeplin}
  G.~J.~Alner {\it et al.},
  Astropart.\ Phys.\  {\bf 28}, 287 (2007)
  [arXiv:astro-ph/0701858];
  G.~J.~Alner {\it et al.}  [UK Dark Matter Collaboration],
  Astropart.\ Phys.\  {\bf 23}, 444 (2005).

\bibitem{edelweiss}
  V.~Sanglard {\it et al.}  [The EDELWEISS Collaboration],
  Phys.\ Rev.\ D {\bf 71}, 122002 (2005)
  [arXiv:astro-ph/0503265].


\bibitem{cresst}
  G.~Angloher {\it et al.},
  Astropart.\ Phys.\  {\bf 23}, 325 (2005)
  [arXiv:astro-ph/0408006].



\bibitem{warp}
  P.~Benetti {\it et al.},
  arXiv:astro-ph/0701286;
  R.~Brunetti {\it et al.},
  New Astron.\ Rev.\  {\bf 49}, 265 (2005)
  [arXiv:astro-ph/0405342].

\bibitem{coupp}
  W.~J.~Bolte {\it et al.},
  J.\ Phys.\ Conf.\ Ser.\  {\bf 39}, 126 (2006).

\bibitem{GGTurner}
  E.~I.~Gates, G.~Gyuk and M.~S.~Turner,
  Phys.\ Rev.\ D {\bf 53} (1996) 4138
  [arXiv:astro-ph/9508071],
 E.~Gates, G.~Gyuk and M.~S.~Turner,
  arXiv:astro-ph/9704253.

\bibitem{Drukier}
  A.~K.~Drukier, K.~Freese and D.~N.~Spergel,
  Phys.\ Rev.\ D {\bf 33}, 3495 (1986).





\bibitem{white}
  A.~Helmi, S.~D.~M.~White and V.~Springel,
  Phys.\ Rev.\ D {\bf 66}, 063502 (2002)
  [arXiv:astro-ph/0201289].



\bibitem{jungman}
  G.~Jungman, M.~Kamionkowski and K.~Griest,
  Phys.\ Rept.\  {\bf 267}, 195 (1996).


\bibitem{KaplanNelson}
  A.~E.~Nelson and D.~B.~Kaplan,
  Phys.\ Lett.\ B {\bf 192}, 193 (1987);
D.~B.~Kaplan and A.~Manohar,
  Nucl.\ Phys.\ B {\bf 310}, 527 (1988).

\bibitem{Bottino}
 A.~Bottino, F.~Donato, N.~Fornengo and S.~Scopel,
   ``Size of the neutralino nucleon cross-section in the light of a new
  Astropart.\ Phys.\  {\bf 18}, 205 (2002)
  [arXiv:hep-ph/0111229].

\bibitem{EllisUpdate}
  J.~R.~Ellis, K.~A.~Olive, Y.~Santoso and V.~C.~Spanos,
  Phys.\ Rev.\ D {\bf 71}, 095007 (2005)
  [arXiv:hep-ph/0502001].



\bibitem{UKQCD}
 C.~Michael, C.~McNeile and D.~Hepburn  [UKQCD Collaboration],
  Nucl.\ Phys.\ Proc.\ Suppl.\  {\bf 106}, 293 (2002)
  [arXiv:hep-lat/0109028].

\bibitem{Weise}
 M.~Procura, T.~R.~Hemmert and W.~Weise,
  Phys.\ Rev.\ D {\bf 69}, 034505 (2004)
  [arXiv:hep-lat/0309020].



\bibitem{scatteraq}
G.~B.~Gelmini, P.~Gondolo and E.~Roulet,
Nucl.\ Phys.\ B {\bf 351}, 623 (1991);
M.~Srednicki and R.~Watkins,
Phys.\ Lett.\ B {\bf 225}, 140 (1989);
M.~Drees and M.~Nojiri,
Phys.\ Rev.\ D {\bf 48}, 3483 (1993)
[arXiv:hep-ph/9307208];
M.~Drees and M.~M.~Nojiri,
Phys.\ Rev.\ D {\bf 47}, 4226 (1993)
[arXiv:hep-ph/9210272];
J.~R.~Ellis, A.~Ferstl and K.~A.~Olive, 
Phys.~Lett.~B  481, (2000) 304,
[arXiv:hep-ph/0001005].

\bibitem{Green:2007rb}
  A.~M.~Green,
  JCAP {\bf 0708}, 022 (2007)
  [arXiv:hep-ph/0703217].

\bibitem{LewinSmith}
 J.~D.~Lewin and P.~F.~Smith,
  Astropart.\ Phys.\  {\bf 6}, 87 (1996).




\bibitem{bringmann}
  T.~Bringmann, L.~Bergstrom and J.~Edsjo,
  JHEP {\bf 0801}, 049 (2008)
  [arXiv:0710.3169 [hep-ph]].

\bibitem{gchist}
  L.~Bergstrom, P.~Ullio and J.~H.~Buckley,
  Astropart.\ Phys.\  {\bf 9}, 137 (1998)
  [arXiv:astro-ph/9712318];
  L.~Bergstrom, J.~Edsjo and P.~Ullio,
  Phys.\ Rev.\ Lett.\  {\bf 87}, 251301 (2001)
  [arXiv:astro-ph/0105048];
  V.~Berezinsky, A.~Bottino and G.~Mignola,
  Phys.\ Lett.\ B {\bf 325}, 136 (1994)
  [arXiv:hep-ph/9402215];
  A.~Cesarini, F.~Fucito, A.~Lionetto, A.~Morselli and P.~Ullio,
  Astropart.\ Phys.\  {\bf 21}, 267 (2004)
  [arXiv:astro-ph/0305075];
  P.~Ullio, L.~Bergstrom, J.~Edsjo and C.~G.~Lacey,
  Phys.\ Rev.\ D {\bf 66}, 123502 (2002)
  [arXiv:astro-ph/0207125].



\bibitem{nfw}
 J.~F.~Navarro, C.~S.~Frenk and S.~D.~M.~White,
  Astrophys.\ J.\  {\bf 462}, 563 (1996)
  [arXiv:astro-ph/9508025];
  J.~F.~Navarro, C.~S.~Frenk and S.~D.~M.~White,
  Astrophys.\ J.\  {\bf 490}, 493 (1997).



\bibitem{ac}
  F.~Prada, A.~Klypin, J.~Flix, M.~Martinez and E.~Simonneau,
  arXiv:astro-ph/0401512;
  G.~Bertone and D.~Merritt,
  Mod.\ Phys.\ Lett.\ A {\bf 20}, 1021 (2005)
  [arXiv:astro-ph/0504422];
  G.~Bertone and D.~Merritt,
  Phys.\ Rev.\ D {\bf 72}, 103502 (2005)
  [arXiv:astro-ph/0501555].

\bibitem{hess}
F.~Aharonian {\it et al.}  [The HESS Collaboration],
  arXiv:astro-ph/0408145;
For example, see F.~Ahronian, Talk at TeV Particle Astrophysics Workshop, Batavia, USA, July 2005. 

\bibitem{magic}
  J.~Albert {\it et al.}  [MAGIC Collaboration],
  Astrophys.\ J.\  {\bf 638}, L101 (2006)
  [arXiv:astro-ph/0512469].

\bibitem{whipple}
  K.~Kosack {\it et al.}  [The VERITAS Collaboration],
  Astrophys.\ J.\  {\bf 608}, L97 (2004)
  [arXiv:astro-ph/0403422].

\bibitem{cangaroo}
 K.~Tsuchiya {\it et al.}  [CANGAROO-II Collaboration],
  Astrophys.\ J.\  {\bf 606}, L115 (2004)
  [arXiv:astro-ph/0403592].



\bibitem{actdark}
  D.~Hooper and J.~March-Russell,
  Phys.\ Lett.\ B {\bf 608}, 17 (2005)
  [arXiv:hep-ph/0412048];
  D.~Hooper, I.~de la Calle Perez, J.~Silk, F.~Ferrer and S.~Sarkar,
  JCAP {\bf 0409}, 002 (2004)
  [arXiv:astro-ph/0404205];
S.~Profumo,
  Phys.\ Rev.\  D {\bf 72}, 103521 (2005)
  [arXiv:astro-ph/0508628];
  L.~Bergstrom, T.~Bringmann, M.~Eriksson and M.~Gustafsson,
  Phys.\ Rev.\ Lett.\  {\bf 94}, 131301 (2005)
  [arXiv:astro-ph/0410359].



\bibitem{hessastro}
  F.~Aharonian and A.~Neronov,
  Astrophys.\ J.\  {\bf 619}, 306 (2005)
  [arXiv:astro-ph/0408303];
  arXiv:astro-ph/0503354;
  AIP Conf.\ Proc.\  {\bf 745}, 409 (2005);
  A.~Atoyan and C.~D.~Dermer,
  Astrophys.\ J.\  {\bf 617}, L123 (2004)
  [arXiv:astro-ph/0410243].



\bibitem{gabi}
  G.~Zaharijas and D.~Hooper,
  Phys.\ Rev.\  D {\bf 73}, 103501 (2006).
  [arXiv:astro-ph/0603540].


\bibitem{Dodelson:2007gd}
  S.~Dodelson, D.~Hooper and P.~D.~Serpico,
  arXiv:0711.4621 [astro-ph].


\bibitem{Evans:2003sc}
  N.~W.~Evans, F.~Ferrer and S.~Sarkar,
  Phys.\ Rev.\  D {\bf 69}, 123501 (2004)
  [arXiv:astro-ph/0311145].

\bibitem{Bergstrom:2005qk}
  L.~Bergstrom and D.~Hooper,
  Phys.\ Rev.\  D {\bf 73}, 063510 (2006)
  [arXiv:hep-ph/0512317].

\bibitem{Strigari:2007at}
  L.~E.~Strigari, S.~M.~Koushiappas, J.~S.~Bullock, M.~Kaplinghat, J.~D.~Simon, M.~Geha and B.~Willman,
  arXiv:0709.1510 [astro-ph].


\bibitem{egdiffuse}
  L.~Bergstrom, J.~Edsjo and P.~Ullio,
  Phys.\ Rev.\ Lett.\  {\bf 87}, 251301 (2001)
  [arXiv:astro-ph/0105048];
  P.~Ullio, L.~Bergstrom, J.~Edsjo and C.~G.~Lacey,
  Phys.\ Rev.\  D {\bf 66}, 123502 (2002)
  [arXiv:astro-ph/0207125];
  D.~Elsaesser and K.~Mannheim,
  Astropart.\ Phys.\  {\bf 22}, 65 (2004)
  [arXiv:astro-ph/0405347];
  D.~Hooper and P.~D.~Serpico,
  JCAP {\bf 0706}, 013 (2007)
  [arXiv:astro-ph/0702328].


\bibitem{glast}
  N.~Gehrels and P.~Michelson,
  Astropart.\ Phys.\  {\bf 11}, 277 (1999);
  S.~Peirani, R.~Mohayaee and J.~A.~de Freitas Pacheco,
  Phys.\ Rev.\ D {\bf 70}, 043503 (2004)
  [arXiv:astro-ph/0401378];
  A.~Cesarini, F.~Fucito, A.~Lionetto, A.~Morselli and P.~Ullio,
  Astropart.\ Phys.\  {\bf 21}, 267 (2004)
  [arXiv:astro-ph/0305075].




\bibitem{pamela}
  A.~Morselli and P.~Picozza,
{\it Prepared for 4th International Workshop on the Identification of Dark Matter (IDM 2002), York, England, 2-6 Sep 2002}.



\bibitem{ams02}
  M.~Sapinski  [AMS Collaboration],
  Acta Phys.\ Polon.\ B {\bf 37}, 1991 (2006);
  C.~Goy  [AMS Collaboration],
  J.\ Phys.\ Conf.\ Ser.\  {\bf 39}, 185 (2006).


\bibitem{baltzpos}
E.~A.~Baltz and J.~Edsjo,
Phys.\ Rev.\ D {\bf 59} (1999) 023511
[arXiv:astro-ph/9808243].




\bibitem{heat}
S.~W.~Barwick {\it et al.}  [HEAT Collaboration],
Astrophys.\ J.\  {\bf 482}, L191 (1997)
[arXiv:astro-ph/9703192];
S.~Coutu {\it et al.}  [HEAT-pbar Collaboration],
in Proceedings of 27th ICRC (2001).

\bibitem{ams01}
Olzem, Jan [AMS Collaboration],
Talk given at the 7th UCLA Symposium on Sources and Detection of Dark Matter and Dark Energy in the Universe, Marina del Ray, CA, Feb 22-24, 2006.



\bibitem{kkpos}
  D.~Hooper and G.~D.~Kribs,
  Phys.\ Rev.\ D {\bf 70}, 115004 (2004)
  [arXiv:hep-ph/0406026].

\bibitem{diffusion}
W.~R.~Webber, M.~A.~Lee and M.~Gupta,
Astrophys.\ J.{\bf 390} (1992) 96;
I.~V.~Moskalenko, A.~W.~Strong, S.~G.~Mashnik and J.~F.~Ormes,
Astrophys.\ J.\  {\bf 586}, 1050 (2003)
[arXiv:astro-ph/0210480];
I.~V.~Moskalenko and A.~W.~Strong,
Phys.\ Rev.\ D {\bf 60}, 063003 (1999)
[arXiv:astro-ph/9905283].


\bibitem{btoc}
D.~Maurin, F.~Donato, R.~Taillet and P.~Salati,
Astrophys.\ J.\  {\bf 555}, 585 (2001)
[arXiv:astro-ph/0101231];
D.~Maurin, R.~Taillet and F.~Donato,
Astron.\ Astrophys.\  {\bf 394}, 1039 (2002)
[arXiv:astro-ph/0206286].




\bibitem{silkpos}
  D.~Hooper and J.~Silk,
  Phys.\ Rev.\ D {\bf 71}, 083503 (2005)
  [arXiv:hep-ph/0409104].

\bibitem{secbg}
  I.~V.~Moskalenko and A.~W.~Strong,
  Astrophys.\ J.\  {\bf 493}, 694 (1998)
  [arXiv:astro-ph/9710124].



\bibitem{hoopertaylorsilk}
  D.~Hooper, J.~E.~Taylor and J.~Silk,
  Phys.\ Rev.\ D {\bf 69}, 103509 (2004)
  [arXiv:hep-ph/0312076].








\bibitem{neutrinoshalo}
  G.~Bertone, E.~Nezri, J.~Orloff and J.~Silk,
  Phys.\ Rev.\  D {\bf 70}, 063503 (2004)
  [arXiv:astro-ph/0403322].



\bibitem{neutrinosun}
 L.~Bergstrom, J.~Edsjo and P.~Gondolo,
  Phys.\ Rev.\  D {\bf 55}, 1765 (1997)
  [arXiv:hep-ph/9607237];
  Phys.\ Rev.\  D {\bf 58}, 103519 (1998)
  [arXiv:hep-ph/9806293];
 V.~D.~Barger, F.~Halzen, D.~Hooper and C.~Kao,
  Phys.\ Rev.\  D {\bf 65}, 075022 (2002)
  [arXiv:hep-ph/0105182];
  V.~Barger, W.~Y.~Keung, G.~Shaughnessy and A.~Tregre,
  Phys.\ Rev.\  D {\bf 76}, 095008 (2007)
  [arXiv:0708.1325 [hep-ph]];
  V.~D.~Barger, W.~Y.~Keung and G.~Shaughnessy,
  arXiv:0709.3301 [astro-ph].

\bibitem{capture}
A.~Gould, Astrophys.\ J.\ {\bf 388}, 338 (1991).


\bibitem{equ1}
K.~Griest and D.~Seckel,
Nucl.\ Phys.\ B {\bf 283}, 681 (1987)
[Erratum-ibid.\ B {\bf 296}, 1034 (1988)].

\bibitem{equ2}
A.~Gould,
Astrophys.\ J.\  {\bf 321}, 571 (1987).

\bibitem{Jungman:1994jr}
G.~Jungman and M.~Kamionkowski,
Phys.\ Rev.\ D {\bf 51} (1995) 328
[arXiv:hep-ph/9407351];
For a more recent calculation see:
 M.~Cirelli, N.~Fornengo, T.~Montaruli, I.~Sokalski, A.~Strumia and F.~Vissani,
  arXiv:hep-ph/0506298.


\bibitem{Halzen:2005ar}
  F.~Halzen and D.~Hooper,
  Phys.\ Rev.\  D {\bf 73}, 123507 (2006)
  [arXiv:hep-ph/0510048].


\bibitem{icecube}
T.~DeYoung  [IceCube Collaboration],
  Int.\ J.\ Mod.\ Phys.\ A {\bf 20}, 3160 (2005);
J.~Ahrens {\it et al.}  [The IceCube Collaboration],
  Nucl.\ Phys.\ Proc.\ Suppl.\  {\bf 118}, 388 (2003)
  [arXiv:astro-ph/0209556].

\bibitem{km3}
  P.~Sapienza,
  Nucl.\ Phys.\ Proc.\ Suppl.\  {\bf 145}, 331 (2005).


\bibitem{superk}
  S.~Desai {\it et al.}  [Super-Kamiokande Collaboration],
  Phys.\ Rev.\  D {\bf 70}, 083523 (2004)
  [Erratum-ibid.\  D {\bf 70}, 109901 (2004)]
  [arXiv:hep-ex/0404025].




\bibitem{amanda}
M.~Ackermann {\it et al.}  [AMANDA Collaboration],
  arXiv:astro-ph/0508518.

\bibitem{baksan}
M.~M.~Boliev {\it et al.}, 
Proc. of the Intl. Workshop on Aspects of Dark Matter in Astrophysics and Particle Physics, Heidelberg, Germany, 1996. Edited by H.~V.~Klapdor-Kleingrothaus, Y.~Ramachers. Singapore, World Scientific, 1997.


\bibitem{macro}
M.~Ambrosio {\it et al.}  [MACRO Collaboration],
  Phys.\ Rev.\ D {\bf 60}, 082002 (1999)
  [arXiv:hep-ex/9812020].


\bibitem{antares}
J.~Hossl  [ANTARES Collaboration],
Proc. of the Identification of Dark Matter (IDM), Edinburgh, Scotland, 2004;
J.~Brunner  [ANTARES Collaboration],
  Nucl.\ Phys.\ Proc.\ Suppl.\  {\bf 145}, 323 (2005).


\bibitem{naiad}
 G.~J.~Alner {\it et al.}  [UK Dark Matter Collaboration],
  Phys.\ Lett.\ B {\bf 616}, 17 (2005)
  [arXiv:hep-ex/0504031].


\bibitem{picasso}
  M.~Barnabe-Heider {\it et al.}  [PICASSO Collaboration],
  arXiv:hep-ex/0502028.





\bibitem{syndm}
 E.~A.~Baltz and L.~Wai,
  Phys.\ Rev.\ D {\bf 70}, 023512 (2004)
  [arXiv:astro-ph/0403528];
  S.~Colafrancesco, S.~Profumo and P.~Ullio,
  Astron.\ Astrophys.\  {\bf 455}, 21 (2006)
  [arXiv:astro-ph/0507575];
  L.~Bergstrom, M.~Fairbairn and L.~Pieri,
  Phys.\ Rev.\  D {\bf 74}, 123515 (2006)
  [arXiv:astro-ph/0607327].


\bibitem{haze}
  D.~P.~Finkbeiner,
  arXiv:astro-ph/0409027;
  D.~Hooper, D.~P.~Finkbeiner and G.~Dobler,
  Phys.\ Rev.\  D {\bf 76}, 083012 (2007),
  arXiv:0705.3655 [astro-ph].










\bibitem{sqgl}
  S.~Dawson, E.~Eichten and C.~Quigg,
  Phys.\ Rev.\ D {\bf 31}, 1581 (1985).



\bibitem{Tovey}
 D.~R.~Tovey,
  Eur.\ Phys.\ J.\ direct C {\bf 4}, N4 (2002).





\bibitem{recon}
  H.~Bachacou, I.~Hinchliffe and F.~E.~Paige,
  Phys.\ Rev.\ D {\bf 62}, 015009 (2000)
  [arXiv:hep-ph/9907518].


 \bibitem{drees}
  M.~Drees, Y.~G.~Kim, M.~M.~Nojiri, D.~Toya, K.~Hasuko and T.~Kobayashi,
  Phys.\ Rev.\ D {\bf 63}, 035008 (2001)
  [arXiv:hep-ph/0007202].

\bibitem{fitter}
  R.~Lafaye, T.~Plehn and D.~Zerwas,
  arXiv:hep-ph/0404282;
 P.~Bechtle, K.~Desch and P.~Wienemann,
  Comput.\ Phys.\ Commun.\  {\bf 174}, 47 (2006)
  [arXiv:hep-ph/0412012].


\bibitem{lhc}
 H.~Bachacou, I.~Hinchliffe and F.~E.~Paige,
  Phys.\ Rev.\ D {\bf 62}, 015009 (2000)
  [arXiv:hep-ph/9907518];
  B.~C.~Allanach, C.~G.~Lester, M.~A.~Parker and B.~R.~Webber,
  JHEP {\bf 0009}, 004 (2000)
  [arXiv:hep-ph/0007009];
  H.~Baer, C.~h.~Chen, F.~Paige and X.~Tata,
  Phys.\ Rev.\ D {\bf 52}, 2746 (1995)
  [arXiv:hep-ph/9503271];
Phys.\ Rev.\ D {\bf 53}, 6241 (1996)
  [arXiv:hep-ph/9512383];
  S.~Abdullin and F.~Charles,
  Nucl.\ Phys.\ B {\bf 547}, 60 (1999)
  [arXiv:hep-ph/9811402].




\bibitem{ditau}
  S.~Abdullin {\it et al.},
  Eur.\ Phys.\ J.\ C {\bf 39S2}, 41 (2005);
  A.~Datta, A.~Djouadi, M.~Guchait and F.~Moortgat,
  Nucl.\ Phys.\ B {\bf 681}, 31 (2004)
  [arXiv:hep-ph/0303095].

\bibitem{higgsmeasure}
  R.~Kinnunen, S.~Lehti, F.~Moortgat, A.~Nikitenko and M.~Spira,
  Eur.\ Phys.\ J.\ C {\bf 40N5}, 23 (2005)
  [arXiv:hep-ph/0503075].

 

\bibitem{fermilab}
  S.~Abel {\it et al.}  [SUGRA Working Group Collaboration],
  arXiv:hep-ph/0003154.


\bibitem{recenttevatron}
  D.~Bortoletto  [CDF and D0 Collaborations],
  PoS {\bf HEP2005}, 347 (2006);
  A.~Canepa  [CDF Collaboration],
  arXiv:hep-ex/0603032;
  A.~Anastassov  [CDF and D0 Collaborations],
  PoS {\bf HEP2005}, 326 (2006);
  V.~Abazov  [D0 Collaboration],
  arXiv:hep-ex/0604029.


\bibitem{Baltz:2006fm}
  E.~A.~Baltz, M.~Battaglia, M.~E.~Peskin and T.~Wizansky,
  Phys.\ Rev.\  D {\bf 74}, 103521 (2006)
  [arXiv:hep-ph/0602187].


\bibitem{ilc}
 G.~Weiglein {\it et al.}  [LHC/LC Study Group],
  arXiv:hep-ph/0410364.





\end{thebibliography}
\end{document}